\documentclass[aps,pra,reprint,a4paper,superscriptaddress,showkeys,floatfix]{revtex4-1}
\usepackage[utf8]{inputenc}              

\usepackage[english]{babel}              
                                         
\usepackage[T1]{fontenc}                 

\usepackage{lmodern}                     

\usepackage{microtype}                   
\usepackage{xcolor}                      

\usepackage{graphicx}                    

\usepackage[bf]{subfigure}

\usepackage{amssymb,amsmath,amsfonts}      

\usepackage{mathtools}

\usepackage{braket}                        

\usepackage{tensor}                        

\usepackage[smaller]{cancel}        

\usepackage{upgreek}     
\usepackage{bm}          
\usepackage[colorlinks,citecolor=blue,filecolor=black,linkcolor=red,urlcolor=blue]{hyperref}
\newcommand{\weq}[1][]{\overset{w_{#1}}{=}}    
\newcommand{\Pb}[2]{\left\{#1,#2\right\}_{P}}    
\newcommand{\Co}[2]{\left[#1,#2\right]_{-}}      
\renewcommand{\bar}[1]{\overline{#1}}
\newcommand{\bs}[1]{\boldsymbol{#1}}
\begin{document}

\title{Localization POVMs and intrinsic temporal uncertainty}

\date{\today}

\author{E. R. F. Taillebois}
\email{emile.taillebois@ifgoiano.edu.br}
\affiliation{Instituto Federal Goiano - Campus Avançado Ipameri, 75.780-000, Ipameri, Goiás, Brazil}

\author{A. T. Avelar}
\affiliation{Instituto de Física, Universidade Federal de Goiás, 74.690-900, Goiânia,
Goiás, Brazil}

\begin{abstract}
The causality issues concerning the localization of relativistic quantum systems, as evidenced by Hegerfeld's paradox, are addressed through a proper-time formalism of single-particle operators. Starting from the premise that physical variables associated to the proper-time gauge have a prominent role in the specification of position, since they do not depend on classical parameters connected to an external observer, we obtain a single-particle formalism in which localization is described by explicitly covariant four-vector operators associated with POVM measurements parametrized by the system's proper-time. Among the consequences of this result, we emphasize that physically acceptable states are necessarily associated with the existence of a temporal uncertainty and their proper-time evolution is not subject to the causality violation predicted by Hegerfeldt.
\end{abstract}

\pacs{}

\keywords{Localization, Causality, Hegerfeldt's theorem}

\maketitle

\section{\label{sec:intro}Introduction}

In the recent literature, a variety of scenarios involving the introduction of relativistic effects in quantum information theory (QIT) have been addressed. These include studies in the context of relativistic quantum mechanics (RQM)\cite{Czachor1997, Peres2002, Gingrich2002, Czachor2003, Gonera2004, Peres2004, Caban2005, Saldanha2012, Taillebois2013, Bartlett2005, Kim2005, Landulfo2009, Dunningham2009, Caban2014}, in which the number of particles is fixed, as well as in the context of quantum field theory (QFT) \cite{Fuentes-Schuller2005,Alsing2006,Caban2006,Fuentes2010,Mehri-Dehnavi2011,Downes2011,Palmer2012}. Several of these works were concerned with relativistic effects over the information stored in the spin degrees of freedom of massive systems, since these are ideal models for quantum bits. Despite the multitude of results concerning such systems, their analyses are generally confronted to ambiguities in the definition and interpretation of the relativistic spin concept \cite{Gingrich2002,Czachor2003,Gonera2004,Peres2004,Caban2005,Saldanha2012,Taillebois2013,Caban2013, Palmer2013,Giacomini2019,Taillebois2020} not to mention issues involving the covariance of these definitions. From the perspective of RQM, such ambiguities are directly related to controversies regarding the definition of a unique localization concept for relativistic quantum systems \cite{Pryce1948,Fleming1965,Lorente1974,Aguilar2013,Silva2019}, an issue that underwent extensive discussion in the past without achieving an acceptable consensus.

Besides being an issue for the appropriate definition of relativistic spin, the localization problem in relativistic quantum systems can also affect certain predictions related to quantum information properties, since data-processing is typically performed by procedures that occur in limited regions of space \cite{Caban2014,Palmer2012}. Although there is no consensus on an unambiguous definition of position for this type of system, the so-called Newton-Wigner operator \cite{Newton1949} has a prominent role in literature, since it behaves like a local 3-vector. However, from the point of view of RQM, the self-adjoint nature of this operator allows strictly localized states to be defined, which leads, according to Hegerfeldt's theorem \cite{Hegerfeldt1974,Hegerfeldt1980}, to the violation of relativistic causality.

Hegerfeldt's theorem established that Hilbert's space structure, together with a sign energy restriction, implies that an initially strictly localized state will exhibit an instantaneous dispersion with non-zero detection probabilities over any region of space, thus violating relativistic causality. As a result, the description of position detection procedures in terms of self-adjoint (s.a.) operators will necessarily be inconsistent with relativistic causality. An alternative to these descriptions is to define the localization concept through POVMs, since they do not necessarily lead to strictly localized states \cite{Terno2014,Celeri2016}. However, it should be noted that inconsistencies concerning causality may also occur in such an approach, since Hegerfeldt's theorem was later extended to states with exponentially bounded decay \cite{Hegerfeldt1985}.

In this paper, the investigation about a possible connection between Hegerfeldt's paradox and the existence of a temporal uncertainty for states with a well-defined energy sign is explored by means of an approach based on physical variables parameterized by the system's proper-time. Although one of the main interests in conducting such a study is the subsequent investigation of its consequences over the spin definition and its properties in QIT, the model to be investigated here will be that of a spinless massive free particle. Such a choice is motivated by the need to adequately underpin the notion of localization resulting from the present proposal before considering potential contributions resulting from the spin inclusion. A condensed report of this work is given in \cite{Taillebois2020prl}.

This work is to be developed in the RQM framework, i.e. in a regime which neglects effects related to creation and annihilation of particles. In this context, the concept of single-particle operators is important, since only this type of quantity can be physically interpreted in RQM. For single-particle systems, such operators are given by the direct sum of operators defined over the subspaces with well defined energy sign. It is worth mentioning that, to be a physically acceptable single-particle observable, a s.a. operator must be written as the direct sum of equally s.a. operators acting on the subspaces with well defined energy sign. This definition will be important for the results to be further developed as it will be related with the need to describe the concept of proper-time-parameterized localization by means of POVMs instead of s.a. operators.

In Section \ref{sec:prop} a classical proper-time approach for the description of a spinless massive free particle is presented, and four-vector coordinates related to an intrinsic notion of position parameterized by the system's proper-time are introduced. The definition of such variables allows to proceed with Dirac's quantization method without the need for a previous classical gauge fixing, a procedure that is presented in Section \ref{sec:quant}. The s.a. extensions of the proper-time parameterized position operators and the consequences of Hegerfeldt's paradox over them are also analyzed in Section \ref{sec:quant}. Finally, driven by the results of the prior sections, a localization POVM parametrized by the system's proper-time is introduced in Section \ref{sec:hegr}. The behavior of the proposed POVM with regard to Hegerfeldt's paradox is also investigated in that section, while additional discussions and conclusions are presented in Section \ref{sec:disc}.

The natural system of units, in which $\hbar=1$ and $c=1$,  will be adopted throughout the work, as well as Minkowski's metric with $(-,+,+,+)$ signature. In addition, Pauli's matrices will be denoted by $\sigma^{j}$, while $\sigma^{0}$ will denote the $2\times 2$ identity matrix. 

\section{\label{sec:prop}Proper-time position variables}

Using the \textit{eibein} formalism \cite{Kalashnikova1997}, the classical Lagrangian function associated to a spinless massive relativistic free particle can be written as
\begin{equation}
L(e,\dot{x}) \equiv \frac{\dot{x}^{\mu}\dot{x}_{\mu}}{2e} - \frac{em^{2}}{2}, \label{eq:Lag}
\end{equation}  
where $x^{\mu}$ are the generalized coordinates, $e$ is an \textit{eibein} variable and the dot derivatives are to be taken with respect to an arbitrary parameterization $\tau$ of the particle's wold line. Using Dirac's formalism for singular Lagrangian systems \cite{Gitman1990}, the above function leads to the primary and secondary constraints
\begin{subequations}
\begin{eqnarray}
\phi^{(1)}(\pi_{e}) = \pi_{e} \approx 0, \label{eq:cons1} \\
\phi^{(2)}(\pi) = \frac{1}{2}(\pi^{\mu}\pi_{\mu} + m^2) \approx 0, \label{eq:cons2}
\end{eqnarray}
\end{subequations}
where $\pi^{\mu} = \frac{\dot{x}^{\mu}}{e}$ are the momenta related to the $x^{\mu}$ coordinates and $\pi_{e}$ is the momentum of the \textit{einbein} $e$. These constraints are first class and the system's total Hamiltonian can be written as $H_{T}(e,\pi,\pi_e,v_e) = e\phi^{(2)}(\pi) + \phi^{(1)}(\pi_e)\dot{e}$.

The singular character of \eqref{eq:Lag} is related to the existence of a gauge symmetry associated to the generator $G = \epsilon(\tau)\phi^{(2)} + \dot{\epsilon}(\tau)\phi^{(1)}$, where $\epsilon(\tau)$ is the $\tau$-dependent parameter of the infinitesimal gauge transformation \cite{Castellani1982}. The resulting gauge transformations connect physical states parameterized by the same value of $\tau$ and are related to the reparameterization invariance of the model.

\subsection{Proper-time gauge}

The proper-time gauge can be obtained by introducing the constraint
\begin{equation}
\phi^{(*)}_{1}(x,\pi,\tau) \equiv \frac{x^{\mu}\pi_{\mu}}{m} + \tau \approx 0 \label{eq:GgConst1}
\end{equation}
and the corresponding consistency relationship, which leads to the additional restriction $\phi^{(*)}_{2}(e) = me - 1 \approx 0$. Denoting by $\weq$ and $\weq[*]$ the weak equalities over the first class ($\Gamma_{\hspace{-0.1cm}\mathrm{T}}$) and the gauge ($\Gamma_{\hspace{-0.1cm}*}$) constrained phase-space surfaces, respectively, one gets that
\[dx^{\mu}dx_{\mu} = e^{2}\pi_{\mu}\pi^{\mu}(d\tau)^{2} \weq -(me d\tau)^{2} \weq[*] - (d \tau)^{2},\]
i.e. the adopted gauge allows to identify $\tau$ as the system's proper-time.

The Dirac brackets $\{\cdot,\cdot\}_{*}$ for this gauge can be easily computed and the non-zero brackets comprising coordinates and momenta are given by
\begin{subequations} \label{eq:Dir1}
\begin{eqnarray}
\left\{x^{\mu},x^{\nu}\right\}_{*}  = \frac{J^{\mu\nu}}{m^2}, \\
\left\{x^{\mu},\pi_{\nu}\right\}_{*}  = \tensor{\eta}{^{\mu}_{\nu}} + \frac{\pi^{\mu}\pi_{\nu}}{m^2},
\end{eqnarray}
\end{subequations}
where $J^{\mu\nu} \equiv x^{\mu}\pi^{\nu} - x^{\nu}\pi^{\mu}$. Since the gauge constraint \eqref{eq:GgConst1} is $\tau$ dependent and $\big\{f, H(e,\pi)\big\}_{*} = 0$, the equation of motion for an arbitrary quantity $f(x,e,\pi,\pi_e;\tau)$ will be given by
\begin{equation}
\dot{f} \weq[*] \frac{\partial f}{\partial \tau} + \frac{1}{2m}\Pb{f}{\pi_{\mu}\pi^{\mu}}, \label{eq:eqmot}
\end{equation}
where $\Pb{\cdot}{\cdot}$ denotes the usual Poisson bracket. From this prescription, one have that $\dot{x}^{\mu} \weq[*] \frac{\pi^{\mu}}{m}$, $\dot{e} \weq[*] 0$, $\dot{\pi}_{\mu} \weq[*] 0$ and $\dot{\pi}_{e} \weq[*] 0$, i.e. the evolution of the coordinates and momenta is in agreement with the interpretation of $\tau$ as the system's proper-time.

In order to eliminate the explicit $\tau$ dependence from $\Gamma_{\hspace{-0.1cm}*}$ and, consequently, the undesirable Poisson bracket from \eqref{eq:eqmot}, a canonical transformation with generator given by
\[G_{2}(x,e,\pi^{\prime},\pi_{e}^{\prime};\tau) = x^{\mu}\pi_{\mu}^{\prime} - \frac{(\pi^{\prime\mu}\pi^{\prime}_{\mu} + m^2)}{2m}\tau + e\pi^{\prime}_{e}\]
is introduced in the unconstrained phase-space $\Gamma$. The old variables ($x^{\mu},e,\pi_{\mu},\pi_{e}$) are then connected to the new ones ($x^{\prime\mu},e^{\prime},\pi^{\prime}_{\mu},\pi^{\prime}_{e}$) by means of the relationships $\pi_{\mu} = \pi^{\prime}_{\mu}$, $\pi_{e} = \pi^{\prime}_{e}$, $e^{\prime} = e$ and
\begin{equation}
x^{\prime\mu} = x^{\mu} - \frac{\pi^{\prime\mu}}{m}\tau. \label{eq:xprimeprop}
\end{equation}
In the new coordinate system, the gauge constraints are written as $\phi^{(*)\prime}_{1} = x^{\prime\sigma}\pi^{\prime}_{\sigma}/m \approx 0$ and $\phi^{(*)\prime}_{2} = me^{\prime} - 1 \approx 0$, thus eliminating the explicit $\tau$ dependence. In addition, the equation of motion in $\Gamma_{\hspace{-0.1cm}*}$ for an arbitrary quantity $f^{\prime}(x^{\prime},e^{\prime},\pi^{\prime},\pi_{e}^{\prime}; \tau)$ is simply given by $\dot{f}^{\prime} \weq[*] \frac{\partial f^{\prime}}{\partial \tau}$, from where it follows that $\dot{x}^{\prime\mu} \weq[*] 0$, $\dot{\pi}^{\prime}_{\mu} \weq[*] 0$, $\dot{e}^{\prime}\weq[*] 0$ and $\dot{\pi}_{e}^{\prime} \weq[*] 0$. Since the transformation is canonical, the Dirac brackets in \eqref{eq:Dir1} remain valid for the new variables.

\subsection{\label{sec:PropVar}Proper-time physical variables}

Out of the system variables, only the momenta $\pi_{\mu}$ and $\pi_{e}$ define Dirac's observables, since their Poisson brackets with respect to the constraints are null. On the other hand, the present model must have six independent physical variables that must be gauge invariant and explicitly independent of $\tau$. However, due to the restriction given in \eqref{eq:cons2}, the momenta $\pi^{\mu}$ define only three of these variables, while the moment $\pi_{e}$ is eliminated as a result of \eqref{eq:cons1}.

Using \eqref{eq:GgConst1} and \eqref{eq:xprimeprop}, a definition of three variables explicitly independent of $\tau$ which allows to complete the set of six physical variables for the model may be written as
\begin{equation}
q^{\mu} = x^{\mu} + \frac{\pi^{\mu}\pi_{\lambda}x^{\lambda}}{m^{2}}, \label{eq:Fis1z}
\end{equation}
the restriction to only three independent variables being given by $q^{\mu}\pi_{\mu} = 2x^{\mu}\pi_{\mu}\phi^{(2)}(\pi)/m^{2} \weq 0$. Their gauge invariance is evidenced by the Poisson brackets $\Pb{q^{\mu}}{\phi^{(1)}} =  0$ and $\Pb{q^{\mu}}{\phi^{(2)}} = \frac{2\pi^{\mu}}{m^{2}}\phi^{(2)} \weq 0$.

Using \eqref{eq:Fis1z}, it can be shown that
\begin{subequations} \label{eq:ParPos1}
\begin{eqnarray}
\Pb{q^{\mu}}{q^{\nu}} = \frac{J^{\mu\nu}}{m^{2}}, \\
\Pb{q^{\mu}}{\pi_{\nu}} = \tensor[]{\eta}{^{\mu}_{\nu}} + \frac{\pi^{\mu}\pi_{\nu}}{m^{2}},
\end{eqnarray}
\end{subequations}
where $J^{\mu\nu} \equiv x^{\mu}\pi^{\nu} - x^{\nu}\pi^{\mu} = q^{\mu}\pi^{\nu} - q^{\nu}\pi^{\mu}$. These results, together with the constrained relation $q^{\mu} \weq[*] x^{\mu} - \frac{\pi^{\mu}}{m}\tau$ and the fact that $\dot{q}^{\mu} \weq 0$, allow to associate the $q^{\mu}$ variables defined on $\Gamma$ to the $\Gamma_{\hspace{-0.1cm}*}$ coordinates $x^{\prime\mu}$ given in \eqref{eq:xprimeprop}. Given that, equation \eqref{eq:xprimeprop} allows to find a set of Dirac observables $q^{\mu}(\tau)$ in $\Gamma$ related to the space-time coordinates $x^{\mu}$ when these are restricted to the proper-time gauge surface $\Gamma_{\hspace{-0.1cm}*}$:
\begin{equation}
q^{\mu}(\tau) \equiv q^{\mu} + \frac{\pi^{\mu}}{m}\tau. \nonumber
\end{equation}
From this definition, one has that $\dot{q}^{\mu}(\tau) \weq \pi^{\mu}/m$ and $q^{\mu}(\tau)\pi_{\mu}/m \weq -\tau$, which agrees with the earlier behavior found for $x^{\mu}$ over $\Gamma_{\hspace{-0.1cm}*}$. The Poisson brackets presented in \eqref{eq:ParPos1} remain true if the $q^{\mu}$ variables are replaced by $q^{\mu}(\tau)$ and the relationship $J^{\mu\nu} = q^{\mu}(\tau)\pi^{\nu} - q^{\nu}(\tau)\pi^{\mu}$ is also easily verifiable.

Finally, the following results for the Poisson brackets of the variables $q^{\mu}(\tau)$ and $\pi_{\mu}$ can be verified:
\begin{subequations} \label{eq:PosPar}
\begin{eqnarray}
\Pb{J^{\mu\nu}}{\pi^{\sigma}} & = & 2\eta^{\sigma[\mu}\pi^{\nu]}, \\
\Pb{J^{\mu\nu}}{J^{\sigma\rho}} & = & 4\eta^{[\mu[\rho}J^{\sigma]\nu]}, \\
\Pb{J^{\mu\nu}}{q^{\sigma}(\tau)} & = & 2\eta^{\sigma[\mu}q^{\nu]}(\tau),
\end{eqnarray}
\end{subequations}
where $a^{[\mu}b^{\nu]} = \frac{1}{2}(a^{\mu}b^{\nu} - a^{\nu}b^{\mu})$. The relations in \eqref{eq:PosPar} indicate that it is reasonable to interpret $J^{\mu\nu}$ and $\pi_{\mu}$ respectively as the total angular and linear momenta of the system. As a result, the physical variables $q^{\mu}(\tau)$ can be interpreted as the four-position coordinates of the system and the parameter $\tau$ can be seen as the system's proper-time whenever the physical quantity of interest is of the form $f(q^{\mu},\pi_{\mu};\tau) = f(q^{\mu}(\tau),\pi_{\mu})$ in the unconstrained phase-space $\Gamma$. This implies the possibility to investigate the proper-time gauge through the physical variables $q^{\mu}(\tau)$ and $\pi_{\mu}$ without performing the phase-space restriction to the constrained surface $\Gamma_{\hspace{-0.1cm}*}$.

\subsection{\label{sec:PhysVarStro}Physical variables with strongly null Dirac brackets}

Despite being suitable to describe the system in the proper-time gauge, it is interesting, from the quantum perspective, to replace the physical variables $q^{\mu}(\tau)$ and $\pi_{\mu}$ by equivalent new variables $Q^{\mu}(\tau) \equiv q^{\mu}(\tau) + \alpha_{1} \phi^{(1)} + \beta_{1} \phi^{(2)}$ and $\Pi_{\mu} \equiv \pi_{\mu} + \alpha_{2} \phi^{(1)} + \beta_{2} \phi^{(2)}$ such that
\begin{subequations} \label{eq:CondStr}
\begin{eqnarray}
\Pb{Q^{\mu}(\tau)}{\phi^{(1)}} = \Pb{Q^{\mu}(\tau)}{\phi^{(2)}} = 0, \\
\Pb{Q^{\mu}(\tau)}{\phi^{(*)}_{1,2}} \weq 0, \\ 
\Pb{\Pi_{\mu}}{\phi^{(1)}} = \Pb{\Pi_{\mu}}{\phi^{(1)}} = 0, \\ 
\Pb{\Pi_{\mu}}{\phi^{(*)}_{1,2}} \weq 0.
\end{eqnarray}
\end{subequations}

From the conditions in \eqref{eq:CondStr}, these new variables can be written as
\begin{subequations} \label{eq:NewVar}
\begin{eqnarray} 
Q^{\mu}(\tau) = Q^{\mu} + \frac{\Pi^{\mu}\tau}{m}, \label{eq:NewQt} \\
\Pi_{\mu} = \pi_{\mu}\left(1 + \frac{\phi^{(2)}}{m^{2}}\right), \label{eq:NewP}
\end{eqnarray}
\end{subequations}
where
\begin{equation}
Q^{\mu} = -\frac{J^{\mu\lambda}\Pi_{\lambda}}{m^{2}}. \label{eq:NewQ} \\
\end{equation}
Since these variables differ from the original ones only by a linear combination of first class terms, the physical interpretation adopted at the end of the last section for $q^{\mu}(\tau)$ and $\pi_{\mu}$ can be extended to the new variables in \eqref{eq:NewVar}. It should also be noted that over $\Gamma_{\hspace{-0.1cm}\mathrm{T}}$ the new variables in \eqref{eq:NewVar} respect the same algebraic properties than the variables $q^{\mu}(\tau)$ and $\pi_{\mu}$.

\section{\label{sec:quant}Proper-time quantization}

The quantization process will be carried out according to Dirac's approach \cite{Gitman1990} and employing the so-called group averaging technique \cite{Marolf1995(1),Marolf1995(2),Ashtekar1995,Louko2006}. Since this approach does not restrict phase-space variables by means of gauge constraints, the proper-time perspective will be retrieved through the quantization of the proper-time physical variables presented in \eqref{eq:NewVar}.

The unconstrained variables $x^{\mu}$, $\pi_{\mu}$, $e$ and $\pi_{e}$ are quantized through the correspondence principle and leads to the commutation relations given by

\noindent \begin{minipage}{0.33\linewidth}
\begin{eqnarray*}
\Co{\hat{\pi}_{\mu}}{\hat{\pi}_{\nu}} & = 0, \\
\Co{\hat{\pi}_{e}}{\hat{\pi}_{\nu}} & = 0, \\
\Co{\hat{e}}{\hat{x}^{\mu}} & = 0,
\end{eqnarray*}
\end{minipage}%
\begin{minipage}{0.33\linewidth}
\begin{eqnarray*}
\Co{\hat{x}^{\mu}}{\hat{x}^{\nu}} & = 0, \\
\Co{\hat{\pi}_{e}}{\hat{x}^{\mu}} & = 0, \\
\Co{\hat{e}}{\hat{\pi}_{\mu}} & = 0,
\end{eqnarray*}
\end{minipage}%
\begin{minipage}{0.34\linewidth}
\begin{eqnarray*}
\Co{\hat{x}^{\mu}}{\hat{\pi}_{\nu}} & = & i\tensor{\eta}{^{\mu}_{\nu}}, \\
\Co{\hat{e}}{\hat{\pi}_{e}} & = & i. 
\end{eqnarray*}
\end{minipage}%
\vspace{\baselineskip}
In the momentum representation, these relations lead to the auxiliary Hilbert space $\mathcal{H}_{aux} = L^{2}(\mathbb{R}^{3+1}\times\mathbb{R},d^{4}\pi\,d\pi_{e})$, for which the internal product is written as
\begin{equation}
\begin{aligned}
\hspace{-0.2cm}\left(\phi(\tau),\psi(\tau)\right)_{aux} & \equiv \left(\phi(\tau)|\psi(\tau)\right) \\ & \equiv \int_{\mathbb{R}^{3+1}} \hspace{-0.6cm}d^{4}\pi\int_{\mathbb{R}}\hspace{-0.2cm}d\pi_{e} \bar{\phi(\pi,\pi_{e};\tau)}\psi(\pi,\pi_{e};\tau).
\end{aligned} \label{eq:ProdIntAux}
\end{equation}
In this representation, $\hat{\pi}_{\mu}$ and $\hat{\pi}_{e}$ are simply multiplicative operators, while $\hat{x}^{\mu}$ and $\hat{e}$ assume the usual complex derivative form.

In order to adopt the proper-time perspective, the variables $\Pi_{\mu}$ and $Q^{\mu}$ must be quantized. For $\Pi_{\mu}$ this is straightforward since operators $\hat{\pi}_{\mu}$ are commutative and the quantization procedure results in $\hat{\Pi}_{\mu} = \hat{\pi}_{\mu}\left(1 + \frac{\hat{\phi}^{(2)}}{m^{2}}\right)$, where $\hat{\phi}^{(2)} = \frac{1}{2}(\hat{\pi}^{\mu}\hat{\pi}_{\mu} + m^2)$. As for $Q^{\mu}$, a symmetrisation procedure is required. Starting from \eqref{eq:NewQ}, this can be carried out in an unambiguous way as 
\begin{equation}
\hat{Q}^{\mu} \equiv -\frac{\hat{J}^{\mu\lambda}:\hat{\Pi}_{\lambda}}{m^2}, \nonumber
\end{equation}
where $\hat{a}:\hat{b} \equiv \frac{1}{2}(\hat{a}\hat{b} + \hat{b}\hat{a})$.

\subsection{\label{sec:Hphys}Physical Hilbert space}

As part of the process of constructing the so-called physical Hilbert space ($\mathcal{H}_{phys}$), the group averaging technique will be adopted to obtain a gauge invariant internal product starting from the auxiliary internal product \eqref{eq:ProdIntAux}. In the present case, the gauge group is unimodular and the group averaging map for a fixed value of $\tau$ may be defined as
\begin{equation}
\begin{aligned}
\upeta : \mathcal{H}_{aux} & \rightarrow \mathcal{H}^{*}_{aux} \\
            |\phi(\tau)) & \mapsto \frac{m}{(2\uppi)^{2}}(\phi(\tau)|\int_{-\infty}^{\infty}\hspace{-0.4cm}d\alpha\int_{-\infty}^{\infty}\hspace{-0.4cm}d\beta e^{i\alpha \hat{\phi}^{(1)}}e^{i\beta\hat{\phi}^{(2)}}.
\end{aligned} \label{eq:GrAv}
\end{equation} 
The gauge invariance of the Haar measure in \eqref{eq:GrAv} allows to define $\mathcal{H}_{phys}$ as the Hilbert space composed by states
$\upeta(\phi(\tau))$ with a $\tau$ independent internal product given by $(\phi,\psi)_{phys} = \eta(\phi(\tau))[\psi(\tau)]$. Using the momentum completeness relation in $\mathcal{H}_{aux}$, this internal product can be rewritten as
\begin{equation}
(\phi,\psi)_{phys} =  \int_{\mathbb{R}^{3}} d\mu(\bs{\pi}) \phi^{\dagger}(\bs{\pi})\psi(\bs{\pi}), \nonumber
\vspace{\baselineskip}
\end{equation}
where $d\mu(\bs{\pi}) = md^{3}\pi/E_{\bs{\pi}}$ and $\phi(\bs{\pi}) \equiv \left(\phi_{+}(\bs{\pi}) \;\; \phi_{-}(\bs{\pi}) \right)^{T}$, with $\phi_{\xi}(\bs{\pi}) = \phi(\pi^{\prime 0 } = \xi E_{\bs{\pi}}, \bs{\pi}^{\prime} = \xi\bs{\pi}, \pi^{\prime}_{e} = 0)$ and $E_{\bs{\pi}} = \sqrt{\|\bs{\pi}\|^2 + m^{2}}$. It follows that the physical Hilbert space can be written as $\mathcal{H}_{phys} = \mathcal{H}_{phys}^{+} \oplus \mathcal{H}_{phys}^{-} = L^{2}(\mathbb{R}^{3},d\mu(\bs{\pi}))\oplus L^{2}(\mathbb{R}^{3},d\mu(\bs{\pi}))$, with each $L^{2}(\mathbb{R}^{3},d\mu(\bs{\pi}))$ subspace being associated to a well defined energy sign.

\subsection{\label{sec:OpHphys} Four-position s.a. operators over \texorpdfstring{$\mathcal{H}_{phys}$}{Hphys}}

An operator $\hat{A}$ that acts on $\mathcal{H}_{aux}$ and that strongly commutes with the system's constraints may have its behavior restricted to $\mathcal{H}_{phys}$ through the group averaging map $(\phi, \hat{A}_{phys}\psi)_{phys} = \upeta(\phi)[\hat{A}\psi]$ \cite{Ashtekar1995}. This procedure justifies the choice made in Section \ref{sec:PhysVarStro} to replace the variables $q^{\mu}$ by the equivalent variables $Q^{\mu}$ that strongly commute with the system's constraints.

Applying the aforesaid prescription, acting rules for the physical restriction of $\hat{\Pi}_{\mu}$ and $\hat{Q}^{\mu}$ can be obtained. They are given by:
\begin{widetext}
\begin{eqnarray*}
(\phi,\hat{\Pi}_{phys}^{\mu}\psi)_{phys} & = & \int_{\mathbb{R}^{3}}d\mu(\bs{\pi}) \phi^{\dagger}(\bs{\pi})\sigma^{3}\left(E_{\bs{\pi}} \tensor{\eta}{^{\mu}_{0}} + \pi^{j}\tensor{\eta}{^{\mu}_{j}}\right)\psi(\bs{\pi}), \\
(\phi,\hat{Q}_{phys}^{\mu}\psi)_{phys} & = & \int_{\mathbb{R}^{3}}d\mu(\bm{\pi}) \phi^{\dagger}(\bm{\pi})\frac{i\sigma^{3}}{m^2}\left\{\tensor{\eta}{^{\mu}_{0}}E_{\bm{\pi}}\left(\bm{\pi}\cdot\nabla_{\bm{\pi}} + \frac{3}{2}\right) + \tensor{\eta}{^{\mu}_{j}}\left(m^2\frac{\partial}{\partial \pi_{j}} + \pi^{j}\bm{\pi}\cdot\nabla_{\bm{\pi}} + \frac{3}{2}\pi^{j}\right) \right\}\psi(\bm{\pi}).
\end{eqnarray*}
\end{widetext}
These results allow to obtain the quantum analogs over $\mathcal{H}_{phys}$ of the variables $Q^{\mu}(\tau)$, the corresponding operations being given by
\begin{eqnarray*}
\check{Q}_{phys}^{0}(\tau) & = & \sigma^{3}\frac{E_{\bs{\pi}}}{m}\left[\frac{i}{m}\left(\bs{\pi}\cdot\nabla_{\bs{\pi}} + \frac{3}{2}\right) + \tau\right], \\
\check{Q}_{phys}^{j}(\tau) & = & \sigma^{3}\left[i\left(\frac{\partial}{\partial \pi_{j}} + \frac{\pi^{j}}{m^2}\bs{\pi}\cdot\nabla_{\bs{\pi}} + \frac{3}{2}\frac{\pi^{j}}{m^2}\right) +\frac{\pi^{j}}{m}\tau\right].
\end{eqnarray*}
Similarly, the physical restriction of the angular momentum operators can be obtained and correspond to the following operations:
\begin{equation}
\check{J}_{phys}^{0j} = -iE_{\bs{\pi}}\frac{\partial}{\partial \pi_{j}}, \qquad
\check{J}_{phys}^{ij} = i\left(\pi^{j}\frac{\partial}{\partial \pi_{i}} -\pi^{i}\frac{\partial}{\partial \pi_{j}}\right). \nonumber
\end{equation}

According to what was presented in Section \ref{sec:PropVar}, operators $\hat{Q}^{\mu}_{phys}(\tau)$ can be interpreted as the coordinates of the system's four-position parameterized by the proper-time $\tau$. Due to the commutation relations
\begin{equation}
\Co{\hat{J}^{\mu\nu}_{phys}}{\hat{Q}^{\sigma}_{phys}(\tau)} = 2i\eta^{\sigma[\mu}\hat{Q}^{\nu]}_{phys}(\tau), \label{eq:ASD}
\end{equation}
these operators behave like the components of a four-vector quantity with respect to Lorentz transformations, ensuring that they are related to a covariant notion of localization. Besides \eqref{eq:ASD}, it is worth noting that the following commutation relationships are also valid:
\begin{eqnarray*}
\Co{\hat{Q}^{\mu}_{phys}(\tau)}{\hat{Q}^{\nu}_{phys}(\tau)} & = & i\frac{\hat{J}^{\mu\nu}_{phys}}{m^2}, \\
\Co{\hat{J}^{\mu\nu}_{phys}}{\hat{J}^{\sigma\rho}_{phys}} & = & 4i\eta^{[\mu[\rho}\hat{J}^{\sigma]\nu]}_{phys}, \\
\Co{\hat{Q}^{\mu}_{phys}(\tau)}{\hat{\Pi}^{\nu}_{phys}} & = & i\left(\tensor[]{\eta}{^{\mu\nu}} + \frac{\hat{\Pi}^{\mu}_{phys}\hat{\Pi}^{\nu}_{phys}}{m^2}\right), \\
\Co{\hat{J}^{\mu\nu}_{phys}}{\hat{\Pi}^{\sigma}_{phys}} & = & 2i\eta^{\sigma[\mu}\hat{\Pi}^{\nu]}_{phys}.
\end{eqnarray*}
These results show that the operators $\hat{Q}^{\mu}_{phys}(\tau)$ and $\hat{J}^{\mu\nu}_{phys}$ satisfy a deSitter-like algebra with fundamental length given by $1/m$ \cite{Snyder1947,Aldrovandi2016}, the usual canonical behavior of position and momentum variables being recovered only in a instantaneous rest frame or in the non-relativistic limit.

The statement of the s.a. character of the operators $\hat{\Pi}^{\mu}_{phys}$ defined over their natural domain $D_{\Pi_{\mu}} = \left\{\phi|\phi(\bs{\pi}), \sigma^{3}\left(E_{\bs{\pi}} \tensor{\eta}{^{\mu}_{0}} + \pi^{j}\tensor{\eta}{^{\mu}_{j}}\right)\phi(\bs{\pi}) \in \mathcal{H}_{phys}\right\}$ is immediate, since these operators correspond to the direct sum of multiplicative operations. On the other hand, in the case of operators $\hat{Q}^{\mu}_{phys}(\tau)$, it is necessary to verify the s.a. character by calculating the so-called deficiency indices \cite{Gitman2012}, since these will generally be non-limited operators. The following sections will present an analysis concerning the deficiency indices and s.a. extensions of $\hat{Q}^{0}_{phys}(\tau)$ and $\hat{Q}^{3}_{phys}(\tau)$, as well as their complete set of commuting operators (CSCO).

\subsubsection{\label{sec:saq0} Self-adjoint extensions of \texorpdfstring{$\hat{Q}^{0}_{phys}(\tau)$}{Q0phys}}

Together, the operators $\hat{Q}^{0}_{phys}(\tau)$, $\hat{J}^{12}_{phys}$ and $\|\hat{\mathbf{J}}_{phys}\|^{2}$ form a CSCO. For $\hat{J}^{12}_{phys}$ and $\|\hat{J}_{phys}\|^{2}$, the s.a. character is verifiable following the usual non-relativistic approach, the spectrum and the corresponding eigenstates being the same as in the non-relativistic case. On the other hand, for $\hat{Q}^{0}_{phys}(\tau)$, the complete definition of the operator requires a detailed analysis of its deficiency indices and domain.

Using spherical coordinates, it can be shown that the differential operation $\check{Q}^{0}_{phys}(\tau)$ is Lagrange s.a. with regard to the measure $d\mu(\bs{\pi})$, i.e. $(\phi, \check{Q}^{0}_{phys}(\tau)\psi)_{phys} = (\check{Q}^{0}_{phys}(\tau)\phi,\psi)_{phys}, \; \forall \phi,\psi \in C^{\infty}_{0}(\mathbb{R}^3)\oplus C^{\infty}_{0}(\mathbb{R}^3)$. Thus, starting with $C^{\infty}_{0}(\mathbb{R}^3)\oplus C^{\infty}_{0}(\mathbb{R}^3)$ as an initial domain for the definition of the operator $\hat{Q}^{0}_{phys}(\tau)$, the corresponding adjoint $\hat{Q}^{0*}_{phys}(\tau)$ will be completely defined once found its domain $D_{\hat{Q}^{0}_{phys}(\tau)}^{*}$, which is given by
\begin{widetext}
\begin{equation}
D_{\hat{Q}^{0}_{phys}(\tau)}^{*} = \left\{\phi_{*}(r_{\pi})\,|\, r_{\pi}E_{r_{\pi}}\phi_{*}(r_{\pi}) \text{ is a.c. in }\mathbb{R}_{\geq 0}; \; \phi_{*}(r_{\pi}),\check{Q}^{0}_{phys}(\tau)\phi_{*}(r_{\pi}) \in L^{2}(\mathbb{R}_{\geq 0}, d\mu(r_{\pi}))\oplus L^{2}(\mathbb{R}_{\geq 0}, d\mu(r_{\pi}))\right\}, \nonumber
\end{equation} 
\end{widetext}
where  $d\mu(r_{\pi}) = \frac{mr_{\pi}^{2}}{E_{r_{\pi}}}dr_{\pi}$ and a.c. stands for "absolute continuous".

Solving the differential equation $\check{Q}^{0}_{phys}(\tau)R^{t}_{\tau}(r_{\pi}) = tR^{t}_{\tau}(r_{\pi})$ for $t = \pm i/m$, it is found that the solutions 
\begin{equation}
\begin{aligned}
R^{+i/m}_{\tau}(r_{\pi}) \overset{\cdot}{=}
\frac{\sqrt{2}e^{im\tau \ln (\frac{r_{\pi}}{m})}}{r_{\pi}^{1/2}(E_{r_{\pi}} + m)}
\begin{pmatrix}
1 \\
0
\end{pmatrix}, \\
R^{-i/m}_{\tau}(r_{\pi}) \overset{\cdot}{=}
\frac{\sqrt{2}e^{im\tau \ln (\frac{r_{\pi}}{m})}}{r_{\pi}^{1/2}(E_{r_{\pi}} + m)}
\begin{pmatrix}
0 \\
1
\end{pmatrix},
\end{aligned} \nonumber
\end{equation}
belong to $D_{\hat{Q}^{0}_{phys}(\tau)}^{*}$, the first one being associated to the $+ i/m$ eigenvalue while the second one to $-i/m$. This result implies that the operator $\hat{Q}^{0}_{phys}(\tau)$ has deficiency indices $\eta = (1,1)$ and, therefore, possesses an infinite number of s.a. extensions with a single parameter.

It is important to note that, although $\hat{Q}^{0}_{phys}(\tau)$ has s.a. extensions, its projections over the subspaces with well-defined energy signs do not have such extensions, since this projections have deficiency indexes given by $\eta = (1,0)$ (positive energy projection) and $\eta = (0,1)$ (negative energy projection), i.e. these projections are essentially maximally symmetrical operators. As will be seen further on, this implies that a eigenfunction of an s.a. extension of $\hat{Q}^{0}_{phys}(\tau)$ cannot have a well defined energy sign, that is, it will necessarily be formed by a superposition of states with positive and negative energy.

To define the s.a. extensions of the symmetric operator $\hat{Q}^{0}_{phys}(\tau)$ it is first necessary to write its closure $\underline{\hat{Q}^{0}_{phys}(\tau)} = (\hat{Q}^{0*}_{phys}(\tau))^{*} \subseteq \left(\underline{\hat{Q}^{0}_{phys}(\tau)}\right)^{*} = \hat{Q}^{0*}_{phys}(\tau)$. Using the sesquilinear form
\[w_{*}(\phi_{*},\psi_{*}) = (\phi_{*}, \hat{Q}^{0*}_{phys}\psi_{*})_{phys} - (\hat{Q}^{0*}_{phys}\phi_{*}, \psi_{*})_{phys},\]
the closed operator $\underline{\hat{Q}^{0}_{phys}(\tau)}$ is given by the acting rule $\underline{\hat{Q}^{0}_{phys}(\tau)}\underline{\phi} = \hat{Q}^{0*}_{phys}(\tau) \underline{\phi}$ defined over the domain
\begin{equation}
D_{\underline{\hat{Q}^{0}_{phys}(\tau)}} = \left\{ \underline{\phi} \Big|\, \underline{\phi} \in D_{\hat{Q}^{0}_{phys}(\tau)}^{*};\; w_{*}(\underline{\phi},R^{\pm i/m}_{\tau}(r_{\pi})) = 0\right\}, \nonumber
\end{equation}
the requirement $w_{*}(\underline{\phi},R^{\pm i/m}_{\tau}(r_{\pi})) = 0$ being equivalent to the boundary condition
\[\lim_{r_{\pi}\rightarrow\infty}|\underline{\phi_{\xi}(r_{\pi})}| < \lim_{r_{\pi}\rightarrow\infty}\mathcal{O}(r_{\pi}^{-3/2}) = 0.\]

Once defined the closure operator $\underline{\hat{Q}^{0}_{phys}(\tau)}$, the Main Theorem from the theory of s.a. extensions of unbounded operators \cite{Gitman2012} can be used to define the one-parameter family of s.a. extensions $\tensor[_{\varphi}]{\hat{Q}}{^{0}_{\hspace{-0.1cm}phys}}(\tau)$, where $\varphi \in (-\uppi,\uppi]$ denotes the parameter of the s.a. extension.  According to this theorem, the domain of a s.a. extension $\tensor[_{\varphi}]{\hat{Q}}{^{0}_{\hspace{-0.1cm}phys}}(\tau)$ of $\hat{Q}^{0}_{phys}(\tau)$ is given by
\begin{widetext}
\begin{equation}
D_{\hspace{-0.15cm}\tensor[_{\varphi}]{\hat{Q}}{^{0}_{\hspace{-0.1cm}phys}}(\tau)} = \left\{ \phi_{\varphi} \in D_{\hat{Q}^{0}_{phys}(\tau)}^{*} \,\Big|\, \right. w_{*}(R^{+i/m}_{\tau}(r_{\pi}) + e^{i\varphi}R^{-i/m}_{\tau}(r_{\pi}),\phi_{\varphi}) = 0\bigg\}, \nonumber
\end{equation}
\end{widetext}
with $\tensor[_{\varphi}]{\hat{Q}}{^{0}_{\hspace{-0.1cm}phys}}(\tau)\phi_{\varphi} = \hat{Q}^{0*}_{phys}(\tau) \phi_{\varphi}$ as acting rule.  Expanding $w_{*}(R^{+i/m}_{\tau}(r_{\pi}) + e^{i\varphi}R^{-i/m}_{\tau}(r_{\pi}),\phi_{\varphi}) = 0$, this requirement can be shown to be equivalent to the boundary condition
\begin{equation}
\lim_{r_{\pi} \rightarrow \infty} \left[\phi_{\varphi;+}(r_{\pi}) - e^{-i\varphi}\phi_{\varphi;-}(r_{\pi})\right] = 0, \nonumber
\end{equation}
with a decay at infinity that must be bounded according to $|\phi_{\varphi;+}(r_{\pi}) - e^{-i\varphi}\phi_{\varphi;-}(r_{\pi})| < \mathcal{O}(r_{\pi}^{-3/2})$.

The spectrum of the s.a. extension $\tensor[_{\varphi}]{\hat{Q}}{^{0}_{\hspace{-0.1cm}phys}}(\tau)$ does not have a discrete component since there exist no $R^{t}(r_{\pi};\tau) \in L^{2}(\mathbb{R}_{\geq 0},d\mu(r_{\pi}))\oplus L^{2}(\mathbb{R}_{\geq 0},d\mu(r_{\pi}))$ such that $\check{Q}^{0}_{phys}(\tau)R^{t}(r_{\pi};\tau) = tR^{t}(r_{\pi};\tau)$ for $t\in \mathbb{R}$. However, since $(\tensor[_{\varphi}]{\hat{Q}}{^{0}_{\hspace{-0.1cm}phys}}(\tau) -t\hat{I})^{-1}$ exists and is unbounded for all $t \in \mathbb{R}$, this s.a. extension has a continuous spectrum with eigenfunctions given by
\begin{equation}
R^{t}_{\varphi}(r_{\pi},\tau) = \sqrt{\frac{m}{2\uppi}}r_{\pi}^{-3/2}\left(\frac{r_{\pi}}{m}\right)^{im\tau}
\begin{pmatrix}
\left(\frac{r_{\pi}}{E_{r_{\pi}} + m}\right)^{-imt} \\
e^{i\varphi}\left(\frac{r_{\pi}}{E_{r_{\pi}} + m}\right)^{imt}
\end{pmatrix} \nonumber
\end{equation}
for $t \in \mathbb{R}$. The basis formed by these normalized eigenfunctions satisfy the delta orthogonality relation
\begin{equation}
\int_{0}^{\infty}d\mu(r_{\pi})R^{t^{\prime}}(r_{\pi};\tau)^{\dagger}R^{t}(r_{\pi};\tau) = \delta(t-t^{\prime}), \nonumber
\end{equation}
and, as states earlier, do not possess states with a well-defined energy sign. If an orthogonal basis of well-defined energy sign eigenfunctions of $\tensor[_{\varphi}]{\hat{Q}}{^{0}_{\hspace{-0.1cm}phys}}(\tau)$ existed, then it will be possible to write $\tensor[_{\varphi}]{\hat{Q}}{^{0}_{\hspace{-0.1cm}phys}}(\tau)$ as a direct sum $\tensor[_{\varphi}]{\hat{Q}}{^{0+}_{\hspace{-0.1cm}phys}}(\tau) \oplus \tensor[_{\varphi}]{\hat{Q}}{^{0-}_{\hspace{-0.1cm}phys}}(\tau)$, where $\tensor[_{\varphi}]{\hat{Q}}{^{0\pm}_{\hspace{-0.1cm}phys}}(\tau)$ would be s.a. extensions for the projections of $\hat{Q}^{0}_{phys}(\tau)$ over subspaces with well-defined energy sign, a result that would be inconsistent with the deficiency indices found earlier. The impossibility to write $\tensor[_{\varphi}]{\hat{Q}}{^{0}_{\hspace{-0.1cm}phys}}(\tau)$ as a direct sum of s.a. operators implies the nonexistence of single-particle s.a. extension for $\hat{Q}^{0}_{phys}(\tau)$, an important result that will be investigated in more details in Section \ref{sec:hegr}.

It is important to note that, although the eigenfunctions of $\tensor[_{\varphi}]{\hat{Q}}{^{0}_{\hspace{-0.1cm}phys}}(\tau)$ cannot have a well-defined energy sign, states with a well-defined energy sign may belong to the domain of a s.a. extension $\tensor[_{\varphi}]{\hat{Q}}{^{0}_{\hspace{-0.1cm}phys}}(\tau)$ if their non-zero component drop to zero fast enough for $r_{\pi}\rightarrow\infty$. This is consistent with the fact that the domain of the closure is contained in the domain of the s.a. extensions. 

Thus, the complete set of generalized eigenfunctions associated to the CSCO formed by $\tensor[_{\varphi}]{\hat{Q}}{^{0}_{\hspace{-0.1cm}phys}}(\tau)$, $\hat{J}^{12}_{phys}$ and $\|\hat{\mathbf{J}}_{phys}\|^{2}$ is given by $\psi^{t,l,m_{z}}_{\varphi}(\bs{\pi};\tau) = Y^{l,m_{z}}(\Omega_{\pi})R^{t}_{\varphi}(r;\tau)$, where $Y^{l,m_{z}}(\Omega_{\pi})$ are the spherical harmonics with $l=0,1,2,\dots$ and $-l \leq m_{z} \leq l$. These states form a orthogonal basis with orthogonality and completeness relations respectively given by
\begin{widetext}
\begin{eqnarray*}
(\psi^{t^{\prime},l^{\prime},m_{z}^{\prime}}(\tau), \psi^{t,l,m_{z}}(\tau))_{phys} & = & \delta_{l^{\prime}l}\delta_{m_{z}^{\prime}m_{z}}\delta(t^{\prime}-t), \\
\sum_{l=0}^{\infty}\sum_{m_{z}=-l}^{l}\int_{-\infty}^{\infty}dt \psi^{t,l,m_{z}}(\bs{\pi};\tau)\psi^{t,l,m_{z}}(\bs{\pi}^{\prime};\tau)^{\dagger} & = & \frac{E_{\bs{\pi}}}{m}\delta(\bs{\pi}-\bs{\pi}^{\prime})\sigma^{0}.
\end{eqnarray*}
\end{widetext}

\subsubsection{Self-adjoint extensions of \texorpdfstring{$\hat{Q}^{3}_{phys}(\tau)$}{Q3phys}}

A CSCO for $\hat{Q}^{3}_{phys}(\tau)$ can be obtained using the operators $\hat{J}^{12}_{phys}$ and $\hat{O}_{phys} \equiv (\hat{J}_{phys}^{12})^{2} - (\hat{J}_{phys}^{01})^2 - (\hat{J}_{phys}^{02})^2$. To obtain the full description of these operators and investigate their s.a. character, it is useful to adopt the hyperbolic coordinates $\omega_{\pi} \in [0,\infty)$, $\nu_{\pi} \in (-\uppi/2,\uppi/2)$ and $\varphi_{\uppi} \in [0,2\uppi)$, related to the original Cartesian coordinates through
\begin{equation}
\begin{aligned}
\pi^{1} & = m\sinh(\omega_{\pi})\sec(\nu_{\pi})\cos(\varphi_{\pi}), \\
\pi^{2} & = m\sinh(\omega_{\pi})\sec(\nu_{\pi})\sin(\varphi_{\pi}), \\
\pi^{3} & = m\tan(\nu_{\pi}).
\end{aligned} \nonumber
\end{equation}

Using the hyperbolic coordinates, the acting rule of $\hat{Q}^{3}_{phys}(\tau)$ may be written as
\begin{equation}
\check{Q}^{3}_{phys}(\tau) = \left[\frac{i}{m}\left(\frac{\partial}{\partial \nu_{\pi}} + \frac{3}{2}\tan(\nu_{\pi}) \right) + \tan(\nu_{\pi})\tau\right]\sigma^{3}, \nonumber
\end{equation}
an operation that is Lagrange s.a. with regard to the measure $d\mu(\nu_{\pi}) = \sec^{3}(\nu_{\pi})d\nu_{\pi}$. Thus, starting with $C_{0}^{\infty}(-\uppi/2,\uppi/2)\oplus C_{0}^{\infty}(-\uppi/2,\uppi/2)$ as an initial domain for the definition of the operator $\hat{Q}^{3}_{phys}(\tau)$, the corresponding adjoint $\hat{Q}^{3*}_{phys}(\tau)$ will be completely defined once found its domain $D^{*}_{\hat{Q}^{3}_{phys}(\tau)}$, which is given by
\begin{widetext}
\begin{eqnarray*}
D^{*}_{\hat{Q}^{3}_{phys}(\tau)} =  \Big\{\phi_{*}(\nu_{\pi}) & \,|\, & \phi_{*}(\nu_{\pi}) \text{ is a.c. in } [-\pi/2,\pi/2]; \\ & & \phi_{*}(\nu_{\pi}),\check{Q}^{3}_{phys}(\tau)\phi_{*}(\nu_{\pi}) \in L^{2}((-\pi/2,\pi/2), d\mu(\nu_{\pi}))\oplus L^{2}((-\pi/2,\pi/2), d\mu(\nu_{\pi})) \Big\}.
\end{eqnarray*}
\end{widetext}

Solving the differential equation $\check{Q}^{3}_{phys}(\tau)\mathcal{V}^{z}_{\tau}(\nu_{\pi}) = z\mathcal{V}^{z}_{\tau}(\nu_{\pi})$ for $z = \pm i/m$, it is found that all the solutions 
\begin{equation}
\begin{aligned}
\mathcal{V}^{\pm i/m}_{(+);\tau}(\nu_{\pi}) & = \frac{1}{\sqrt{\sinh(\uppi)}}\frac{(\sec\nu_{\pi})^{im\tau}}{(\sec\nu_{\pi})^{3/2}}e^{\pm \nu_{\pi}}\begin{pmatrix} 1 \\ 0 \end{pmatrix} \\
\mathcal{V}^{\pm i/m}_{(-);\tau}(\nu_{\pi}) & = \frac{1}{\sqrt{\sinh(\uppi)}}\frac{(\sec\nu_{\pi})^{im\tau}}{(\sec\nu_{\pi})^{3/2}}e^{\mp\nu_{\pi}}\begin{pmatrix} 0 \\ 1 \end{pmatrix}
\end{aligned} \nonumber
\end{equation}
belong to $D^{*}_{\hat{Q}^{3}_{phys}(\tau)}$. Thus, the operator $\hat{Q}^{3}_{phys}(\tau)$ has deficiency indices $\eta = (2,2)$ and possesses an infinite number of s.a. extensions parameterized by four parameters $\varphi = \{\varphi_{n};\; n = 1,2,3,4;\; \varphi_{n} \in (-\uppi,\uppi]\}$. It is interesting to note that, in contrast to the earlier result for $\hat{Q}^{0}_{phys}(\tau)$, the projections of $\hat{Q}^{3}_{phys}(\tau)$ over the subspaces with well-defined energy sign also have s.a. extensions, since their deficiency indices are given by $\eta_{\pm} = (1,1)$.

The closure $\underline{\hat{Q}^{3}_{phys}(\tau)}$ of $\hat{Q}^{3}_{phys}(\tau)$ is given by the acting rule $\underline{\hat{Q}^{3}_{phys}(\tau)}\underline{\phi} = \hat{Q}^{3*}_{phys}(\tau)\underline{\phi}$ defined for the domain
\begin{equation}
D_{\underline{\hat{Q}^{3}_{phys}(\tau)}} = \left\{\underline{\phi} \,|\, \underline{\phi} \in D^{*}_{\hat{Q}^{3}_{phys}(\tau)}; \, w_{*}\left(\underline{\phi},\mathcal{V}^{\pm i/m}_{(\xi);\tau}\right) = 0 \right\}, \nonumber
\end{equation}
the requirement $w_{*}\left(\underline{\phi},\mathcal{V}^{\pm i/m}_{(\xi);\tau}\right) = 0$ being equivalent to the boundary condition
\begin{equation}
\lim_{\nu_{\pi}\rightarrow \pm \frac{\pi}{2}}\underline{\phi_{\xi}(\nu_{\pi})} = 0 \label{eq:eqh}
\end{equation}
with a decay bounded by $\mathcal{O}((\sec\nu_{\pi})^{-3/2})$ for $\nu_{\pi}\rightarrow\pm\uppi/2$.

Once defined $\underline{\hat{Q}^{3}_{phys}(\tau)}$, the Main Theorem can be used to define the parameterized s.a. extensions of $\hat{Q}^{3}_{phys}(\tau)$ as
\begin{widetext}
\vspace{-\baselineskip}
\begin{equation}
\tensor[_{\varphi}]{\hat{Q}}{^{3}_{\hspace{-0.1cm}phys}}(\tau)\,:\, \left\{
\begin{aligned}
D_{\tensor[_{\varphi}]{\hat{Q}}{^{3}_{\hspace{-0.1cm}phys}}(\tau)} & = \Big\{\phi_{\varphi} \in D^{*}_{\hat{Q}^{3}_{phys}(\tau)} \,\Big|\, w_{*}\left(\phi_{i/m} + \hat{U}(\varphi)\phi_{i/m}, \phi_{\varphi}\right) = 0, \;\forall \phi_{i/m} \in \mathfrak{N}_{-i/m}\Big\} \\
\tensor[_{\varphi}]{\hat{Q}}{^{3}_{\hspace{-0.1cm}phys}}(\tau)&\phi_{\varphi} = \underline{\hat{Q}^{3}_{phys}(\tau)}\underline{\phi} + \frac{i}{m}\phi_{i/m} - \frac{i}{m}\hat{U}(\varphi)\phi_{i/m} = \hat{Q}^{3*}_{phys}(\tau)\phi_{\varphi}
\end{aligned}\right., \nonumber
\end{equation}
\end{widetext}
where $\mathfrak{N}_{\mp i/m}$ denotes the space spanned by $\mathcal{V}^{\pm i/m}_{(\xi);\tau}(\nu_{\pi})$, with $\xi=\pm$, and the isometric map $\hat{U}(\varphi)$ is given by
\begin{equation}
\begin{aligned}
\hat{U}(\varphi) : \mathfrak{N}_{-i/m} & \rightarrow \mathfrak{N}_{+i/m} \\
\mathcal{V}^{i/m}_{(\xi);\tau}(\nu_{\pi}) & \mapsto \hat{U}(\varphi)\mathcal{V}^{i/m}_{(\xi);\tau}(\nu_{\pi}) = \hspace{-1pt}\sum_{\xi^{\prime}}U_{\xi^{\prime}\xi}(\varphi)\mathcal{V}^{-i/m}_{(\xi^{\prime});\tau}(\nu_{\pi})
\end{aligned} \nonumber
\end{equation}
with factors $U_{\xi^{\prime}\xi}(\varphi)$ forming an arbitrary $U(2)$ matrix:
\begin{equation}
U(\varphi) \;\dot{=}\; e^{i\varphi_{1}}
\begin{pmatrix}
e^{i\varphi_{2}}\cos(\varphi_{4})   & e^{i\varphi_{3}}\sin(\varphi_{4})  \\
-e^{-i\varphi_{3}}\sin(\varphi_{4}) & e^{-i\varphi_{2}}\cos(\varphi_{4})
\end{pmatrix}.\label{eq:Mat}
\end{equation}

The form of the matrix in \eqref{eq:Mat} shows that the s.a. extensions $\tensor[_{\varphi}]{\hat{Q}}{^{3}_{\hspace{-0.1cm}phys}}(\tau)$ usually do not define single-particle observables, since $\hat{U}(\varphi)$ generally mixes components with distinct energy signs. However, single-particle s.a. extensions can be constructed through the direct sum of the s.a. extensions associated with the projections $\hat{Q}^{3,\pm}_{phys}(\tau)$ over subspaces with well-defined energy sing. These extensions correspond to the operators $\tensor[_{\varphi}]{\hat{Q}}{^{3}_{\hspace{-0.1cm}phys}}(\tau)$ with $\varphi_{2}=\varphi_{4}=0$ and their properties will be analyzed hereafter. 

The well-defined energy sign s.a. extensions $\tensor[_{\varphi}]{\hat{Q}}{^{3,\xi}_{\hspace{-0.4cm}phys}}(\tau)$, with $\xi = \pm$, are given by the acting rule $\tensor[_{\varphi}]{\hat{Q}}{^{3,\xi}_{\hspace{-0.4cm}phys}}(\tau)\phi_{\varphi} = \hat{Q}^{3*,\xi}_{phys}(\tau)\phi_{\varphi}$ with domain $D_{\tensor[_{\varphi}]{\hat{Q}}{^{3,\xi}_{\hspace{-0.4cm}phys}}(\tau)}$ given by the set
\begin{equation}
 \left\{\phi_{\varphi} \in D^{*}_{\hat{Q}^{3,\xi}_{phys}(\tau)} \,|\, w_{*}\left(\mathcal{V}^{i/m}_{(\xi);\tau} + e^{i\varphi}\mathcal{V}^{-i/m}_{(\xi);\tau}, \phi_{\varphi}\right) = 0\right\}, \nonumber
\end{equation}
the restriction imposed by this domain being equivalent to the boundary condition
\begin{widetext}
\begin{equation}
 \lim_{\nu_{\pi}\rightarrow\frac{\pi}{2}}\left[\left(e^{\xi\uppi/2} + e^{-i\varphi}e^{-\xi\uppi/2}\right)\phi_{\varphi}(\nu_{\pi})
 -  \left(e^{-\xi\uppi/2} + e^{-i\varphi}e^{\xi\uppi/2}\right)\phi_{\varphi}(-\nu_{\pi})\right] < \lim_{\nu_{\pi}\rightarrow\frac{\pi}{2}}\mathcal{O}\left((\sec\nu_{\pi})^{-3/2}\right) = 0. \label{eq:Bound1}
\end{equation}
\end{widetext}
Thus, the domain $D_{\hspace{-0.1cm}\tensor[_{\varphi}]{\hat{Q}}{^{3}_{\hspace{-0.1cm}phys}}(\tau)}$ of a single-particle s.a. extension $\tensor[_{\varphi}]{\hat{Q}}{^{3}_{\hspace{-0.1cm}phys}}(\tau)$ will be given by $D_{\tensor[_{\varphi}]{\hat{Q}}{^{3,+}_{\hspace{-0.4cm}phys}}(\tau)}\oplus D_{\tensor[_{\varphi}]{\hat{Q}}{^{3,-}_{\hspace{-0.4cm}phys}}(\tau)}$, its spectrum being fully defined through the spectra of the well-defined energy sign s.a. operators $\tensor[_{\varphi}]{\hat{Q}}{^{3,\pm}_{\hspace{-0.1cm}phys}}(\tau)$.

Before proceeding with the formal spectrum investigation of the operators $\tensor[_{\varphi}]{\hat{Q}}{^{3}_{\hspace{-0.1cm}phys}}(\tau)$, it is worth noting that the internal product of the solutions
\begin{equation}
\mathcal{V}^{z}_{(\xi);\tau}(\nu_{\pi}) = \frac{\mathcal{N}^{z}_{\xi}}{(\sec\nu_{\pi})^{3/2}}e^{im\tau\ln(\sec\nu_{\pi})}e^{-im\xi z \nu_{\pi}}
\begin{pmatrix}
\delta_{\xi +} \\
\delta_{\xi -}
\end{pmatrix} \nonumber
\end{equation}
of $\check{Q}^{3}_{phys}(\tau)\mathcal{V}^{z}_{(\xi);\tau}(\nu_{\pi}) = z\mathcal{V}^{z}_{(\xi);\tau}(\nu_{\pi})$ is given by
\begin{equation}
\begin{aligned}
\int_{-\pi/2}^{\pi/2}d\mu(\nu_{\pi})&\mathcal{V}^{z^{\prime}}_{(\xi^{\prime});\tau}(\nu_{\pi})^{\dagger}\mathcal{V}^{z}_{(\xi);\tau}(\nu_{\pi}) \\ & = \frac{2\delta_{\xi^{\prime}\xi}\bar{\mathcal{N}^{z^{\prime}}_{\xi}}\mathcal{N}^{z}_{\xi}\sin\left(m(z^{\prime} - z)\frac{\uppi}{2}\right)}{m(z^{\prime}-z)}. 
\end{aligned}\label{eq:ProdNO}
\end{equation}
From \eqref{eq:ProdNO} it is evident that, for a fixed energy sign, the orthogonality of the solutions $\mathcal{V}^{z}_{(\xi);\tau}(\nu_{\pi})$ will happen only if $(z^{\prime} - z) = 2n/m$, with $n\in \mathbb{Z}^{*}$. This result shows that the spectrum of the s.a. extensions $\tensor[_{\varphi}]{\hat{Q}}{^{3}_{\hspace{-0.1cm}phys}}(\tau)$ will necessarily be discrete, with normalization constants $\mathcal{N}^{z}_{\xi} = \uppi^{-1/2}$. 

To formally obtain the spectra of operators $\tensor[_{\varphi}]{\hat{Q}}{^{3}_{\hspace{-0.1cm}phys}}(\tau)$ it is necessary to verify if the solutions $\mathcal{V}^{z}_{(\xi);\tau}(\nu_{\pi})$ belong to the domain $D_{\hspace{-0.1cm}\tensor[_{\varphi}]{\hat{Q}}{^{3}_{\hspace{-0.1cm}phys}}(\tau)}$. Since $\mathcal{V}^{z}_{(\xi);\tau}(\nu_{\pi}) \in D^{*}_{\hat{Q}^{3}_{phys}(\tau)}$, it remains to verify the consequences of the boundary condition \eqref{eq:Bound1}. Applying \eqref{eq:Bound1} to $\mathcal{V}^{z}_{(\xi);\tau}(\nu_{\pi})$ it results that $\mathcal{V}^{z}_{(\xi);\tau}(\nu_{\pi}) \in D_{\hspace{-0.1cm}\tensor[_{\varphi}]{\hat{Q}}{^{3}_{\hspace{-0.1cm}phys}}(\tau)}$ only for eigenvalues $z_{n}$ given by
\begin{equation}
z_{\varphi}^{n} = \frac{2}{m\uppi}\left\{\,\mathrm{arctan}\left[\frac{(1-\cos\varphi)}{\sin\varphi}\tanh\left(\frac{\uppi}{2}\right)\right] + n\pi\right\}, \nonumber
\end{equation}
with $n \in \mathbb{Z}$ and $-\pi/2 \leq \mathrm{arctan}(\alpha) \leq \pi/2$, a result that is in agreement with the earlier prevision based on the internal product \eqref{eq:ProdNO} and the orthogonality of the eigenfunctions of s.a. operators. Thus, the eigenfunctions of the s.a. extension $\tensor[_{\varphi}]{\hat{Q}}{^{3}_{\hspace{-0.1cm}phys}}(\tau)$ are given by
\begin{equation}
\mathcal{V}^{z^{n}_{\varphi}}_{(\xi);\tau}(\nu_{\pi}) = \frac{e^{im\tau\ln(\sec\nu_{\pi})}e^{-im\xi z^{n}_{\varphi} \nu_{\pi}}}{\uppi^{1/2}(\sec\nu_{\pi})^{3/2}}\begin{pmatrix}
\delta_{\xi +} \\
\delta_{\xi -}
\end{pmatrix}, \nonumber
\end{equation}
while their orthogonality and completeness relations may be written, respectively, as
\begin{eqnarray*}
\int_{-\infty}^{\infty}d\mu(\nu_{\pi})\mathcal{V}^{z^{n^{\prime}}_{\varphi}}_{(\xi^{\prime});\tau}(\nu_{\pi})^{\dagger}\mathcal{V}^{z^{n}_{\varphi}}_{(\xi);\tau}(\nu_{\pi}) & = & \delta_{\xi^{\prime}\xi}\delta_{n^{\prime}n}, \\
\sum_{\xi}\sum_{n\in\mathbb{Z}}\mathcal{V}^{z^{n}_{\varphi}}_{(\xi);\tau}(\nu_{\pi}^{\prime})\mathcal{V}^{z^{n}_{\varphi}}_{(\xi);\tau}(\nu_{\pi})^{\dagger} & = & \frac{\delta(\nu_{\pi}^{\prime} - \nu_{\pi})}{\sec^{3}\nu_{\pi}}\sigma^{0}.
\end{eqnarray*}

\begin{figure}[h]
\includegraphics[width=0.9\linewidth]{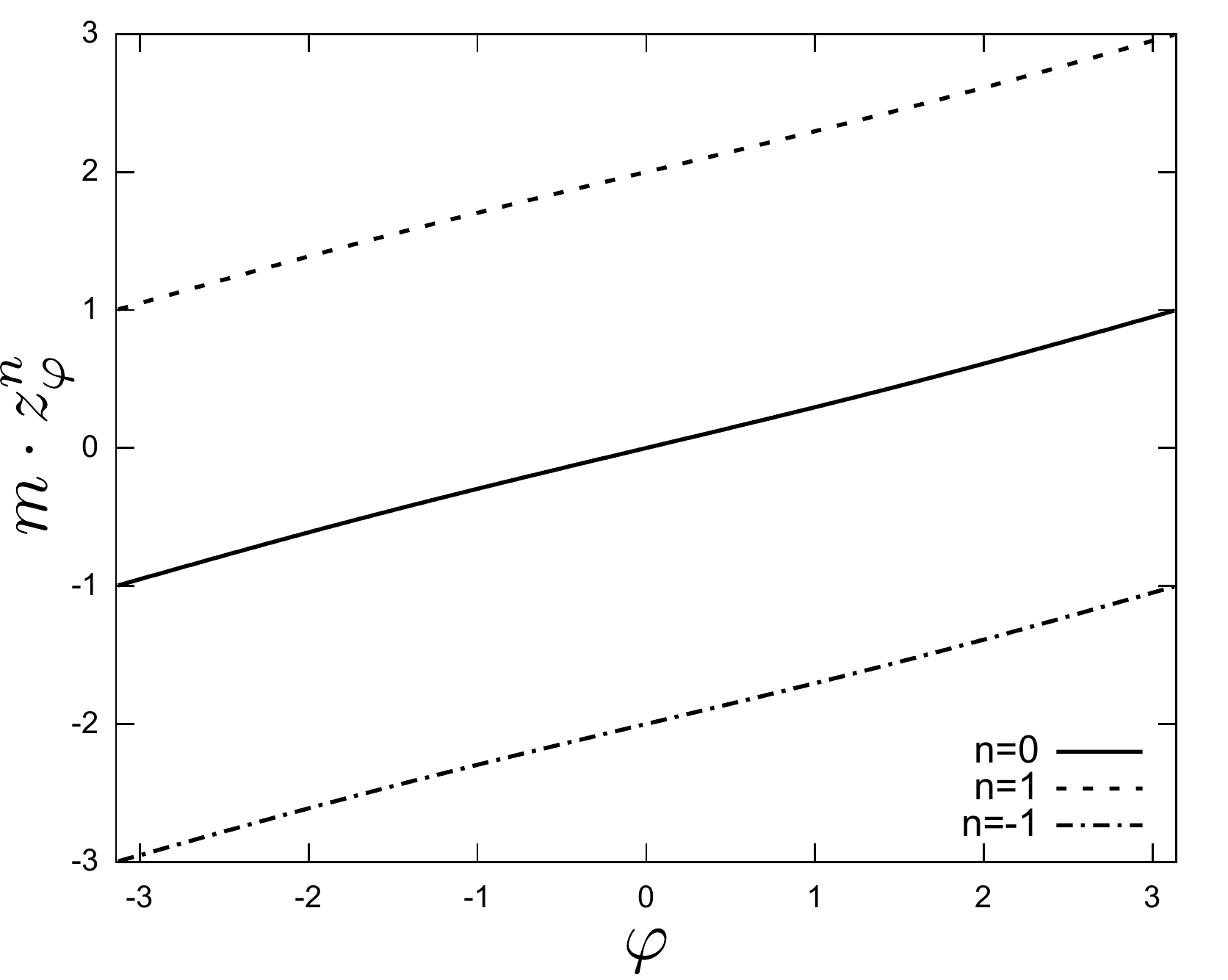}
\caption{\label{fig:zcont} Values assumed by $mz^{n}_{\varphi}$ ($n=0,\pm 1$) as a function of the parameter $\varphi$ that defines the s.a. extensions. For a fixed value of $\varphi$ the eigenvalues  $z^{n}_{\varphi}$ are discrete and separated by increments of $2/m$. However, considering the set of all s.a. extensions $\tensor[_{\varphi}]{\hat{Q}}{^{3}_{\hspace{-0.1cm}phys}}(\tau)$, the spatial continuity is recovered by varying $\varphi$ in the $(-\uppi,\uppi]$ interval.}
\end{figure}

The discrete spectrum of the s.a. extensions $\tensor[_{\varphi}]{\hat{Q}}{^{3}_{\hspace{-0.1cm}phys}}(\tau)$ may seem unsatisfactory at first, since a continuous spectrum is expected for observables associated with the system's position. However, as can be seen in Figure \ref{fig:zcont}, continuity can be recovered when the set of all s.a. extensions $\tensor[_{\varphi}]{\hat{Q}}{^{3}_{\hspace{-0.1cm}phys}}(\tau)$, with $\varphi\in (-\uppi,\uppi]$, is taken into account, since $z_{\uppi}^{n} = \lim_{\varphi \rightarrow -\uppi}z^{n+1}_{\varphi}$. Such a property will be useful later in the construction of POVM measurements related to the system's localization.

To finish the description of the s.a. extensions of $\hat{Q}^{3}_{phys}(\tau)$ it is necessary to verify the properties of the others operators in its CSCO, a task that is presented in Appendix \ref{app:CSCO}. According to the results presented there, the generalized eigenfunctions of the s.a. CSCO of $\tensor[_{\varphi}]{\hat{Q}}{^{3}_{\hspace{-0.1cm}phys}}(\tau)$ are given by
\begin{widetext}
\begin{equation}
\begin{aligned}
\psi^{z^n_{\varphi},\lambda,m_{z}}_{(\xi);\tau}(\bs{\pi}) & = \mathcal{V}^{z^{n}_{\varphi}}_{(\xi);\tau}(\nu_{\pi})\Phi^{m_{z}}(\varphi_{\pi})\mathcal{W}^{\lambda}_{m_{z}}(\omega_{\pi}) \\
& = \frac{\sqrt{\sinh(\uppi\Lambda(\lambda))}}{2}\frac{|\Gamma(\frac{1}{2} + |m_{z}| + i\Lambda(\lambda))|}{(m\uppi)^{3/2}}\frac{e^{im\tau\ln(\sec\nu_{\pi})}e^{-im\xi z^{n}_{\varphi}\nu_{\pi}}e^{im_{z}\varphi_{\pi}}}{(\sec\nu_{\pi})^{3/2}}P^{-|m_{z}|}_{-\frac{1}{2} + i\Lambda(\lambda)}(\cosh\omega_{\pi})\begin{pmatrix}
\delta_{\xi +} \\
\delta_{\xi -}
\end{pmatrix},
\end{aligned} \label{eq:EigenQ3}
\end{equation}
\end{widetext}
where $\Gamma(\cdot)$ is the Gamma function, $P^{-|m_{z}|}_{-\frac{1}{2} + i\Lambda(\lambda)}(\cosh\omega_{\pi})$ are associated conical functions, $\Lambda(\lambda) = \sqrt{-\frac{1}{4}-\lambda}$, $m_z \in \mathbb{Z}$ are the eigenvalues of the s.a. operator $\hat{J}^{12}_{phys}$ and $\lambda \in (-\infty,-\frac{1}{4}]$ are the values of the continuous spectrum of the s.a. extension of $\hat{O}_{phys}$. Thus, the orthogonality and completeness relations of this CSCO are respectively given by
\begin{widetext}
\begin{subequations} \label{eq:Iju}
\begin{eqnarray}
\int_{\mathbb{R}^{3}}d\mu(\bs{\pi})\psi^{z^{n^{\prime}}_{\varphi},\lambda',m_{z}'\dagger}_{(\xi');\tau}(\bs{\pi})\psi^{z^n_{\varphi},\lambda,m_{z}}_{(\xi);\tau}(\bs{\pi}) & = & \delta_{\xi'\xi}\delta_{m_{z}'m_{z}}\delta_{n'n}\delta(\lambda'-\lambda), \\
\sum_{\xi=\pm 1}\sum_{n = -\infty}^{\infty}\sum_{m_{z}=-\infty}^{\infty}\int_{-\infty}^{-1/4}d\lambda \psi^{z^n_{\varphi},\lambda,m_{z}}_{(\xi);\tau}(\bs{\pi}')\psi^{z^n_{\varphi},\lambda,m_{z}\dagger}_{(\xi);\tau}(\bs{\pi}) & = & \frac{E_{\bs{\pi}}}{m}\delta(\bs{\pi}' - \bs{\pi})\sigma^{0}.
\end{eqnarray}
\end{subequations}
\end{widetext}
The physical interpretation of the s.a. extension of the operator $\hat{O}_{phys}$ and its properties will be left to the future, since it is not the main interest of the present work.

\subsection{\label{sec:hegrsa}Hegerfeldt's paradox for the single-particle s.a. extensions of \texorpdfstring{$\hat{Q}^{3}_{phys}(\tau)$}{Q3phys}}

The fact that $\hat{Q}^{3}_{phys}(\tau)$ has s.a. single-particle extensions suggests, at first, that the operators $\tensor[_{\varphi}]{\hat{Q}}{^{j}_{\hspace{-0.1cm}phys}}(\tau)$ should be interpreted as the observables associated with the system's position parameterized by the proper-time. However, there are some considerations that indicate that this interpretation should be refuted, starting with the fact that the discrete spectrum of these operators does not allow a continuous description of the system's position by means of a single s.a. extension with fixed $\varphi$ parameter.

\begin{figure}[b]
\includegraphics[width=0.9\linewidth]{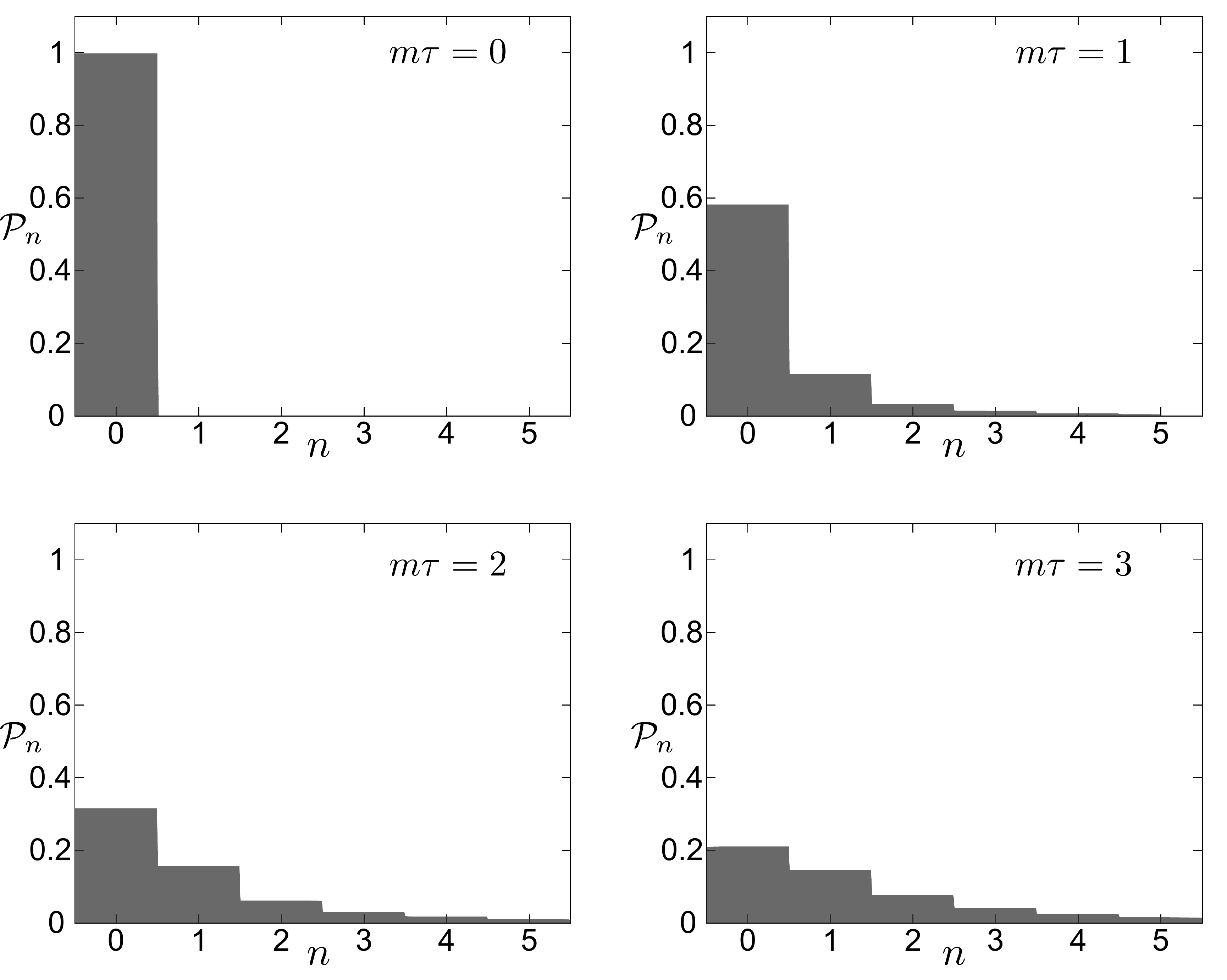}
\caption{\label{Fig:Prob1} Graphical representation of $P_{n}(\tau)$ for $m\tau = 0,1,2$ and $3$. For $m\tau = 0$, the probability is entirely concentrated in $n=0$, which characterizes \eqref{eq:Esta} as a strictly localized state on the z-axis. Causality violation can be observed for other values of $m\tau$, since non-null probabilities are found for all $n > m\tau/2$.}
\end{figure}

A second problem in the description of the position in terms of the s.a. extensions $\tensor[_{\varphi}]{\hat{Q}}{^{j}_{\hspace{-0.1cm}phys}}(\tau)$ is related to Hegerfeldt's paradox. Since the operator $\tensor[_{\varphi}]{\hat{Q}}{^{3}_{\hspace{-0.1cm}phys}}(\tau)$ is s.a., its eigenstates describe strictly localized states on the z-axis and, therefore, must be subject to the causality violation predicted by Hegerfeldt. To verify that, one start with the $\tensor[_{0}]{\hat{Q}}{^{3}_{\hspace{-0.1cm}phys}}(0)$ positive energy eigenstate
\begin{equation}
\ket{\psi} = \int_{-\infty}^{-1/4}d\lambda \alpha(\lambda)\ket{\psi^{z^{0}_{0},\lambda,m_{z}}_{0;+}}. \label{eq:Esta}
\end{equation} 
This state has eigenvalue $z^{0}_{0} = 0$ and, therefore, is strictly localized at the z-axis origin for $\tau = 0$. The probability of finding the state \eqref{eq:Esta} in an eigenvalue $z^{n}_{0} = 2n/m$ ($n\in\mathbb{Z}$) from $\tensor[_{0}]{\hat{Q}}{^{3,+}_{\hspace{-0.4cm}phys}}(\tau)$ for $\tau\neq 0$ is given by
\begin{equation}
\begin{aligned}
P_{n}(\tau) & = \sum_{m_{z}'\in\mathbb{Z}}\int_{-\infty}^{-1/4}d\lambda'|\braket{\psi^{z^{n}_{0},\lambda',m_{z}'}_{\tau;+}|\psi}|^2 \\ & = \frac{m\uppi\tau}{\sinh(m\uppi\tau)}\Big|\mathrm{sinc}\left(\frac{\uppi}{2}(2n-im\tau)\right)\Big|^{2}.
\end{aligned} \nonumber
\end{equation}

\begin{figure}[b]
\centering
\includegraphics[width=0.8\linewidth]{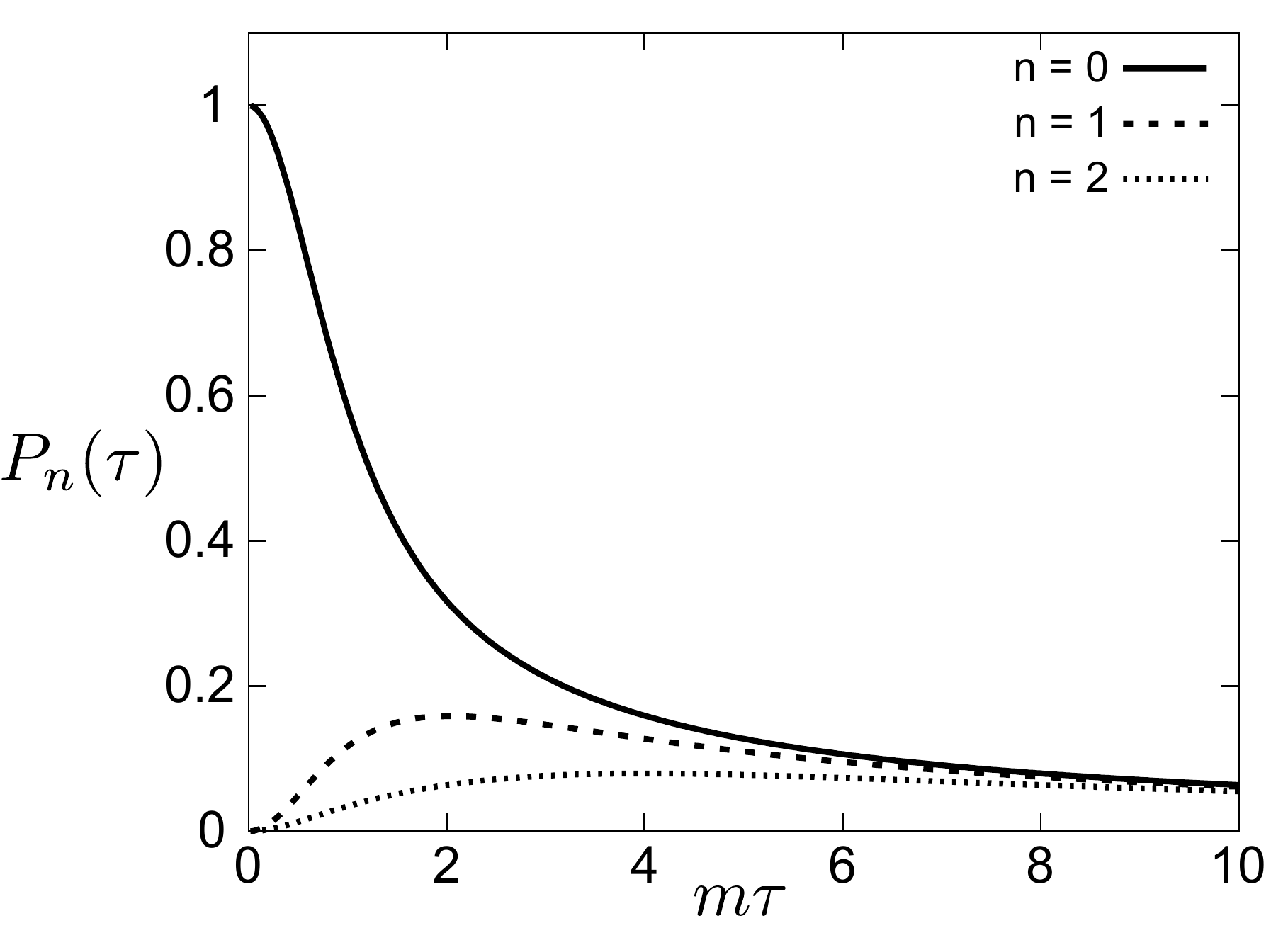}
\caption{\label{Fig:Prob2} Detection probability $P_{n}(\tau)$ as a function of $\tau$ for $n=0,1$ and $2$. Non-null values of $P_{n}(\tau)$ for $n\neq 0$ emerge instantly for $\tau > 0$. This implies that immediately after the proper-time instant $\tau = 0$, where the state is strictly localized, the system has a non-null probability of being detected in $z^{n}_{0}$ for any $n\in\mathbb{Z}$.}
\end{figure}

The quantity $m\tau$ describes the number of Compton wavelengths of the system in a proper-time interval $\tau$. Thus, in order to avoid causality violation for the state \eqref{eq:Esta}, it is necessary that $P_{n}(\tau)$ be null for all $n > m\tau/2$. However, as can be seen in Figure \ref{Fig:Prob1}, the probability $P_{n}(\tau)$ does not respect this condition. Consequently, causality is violated for a description of the system's position in terms of the single-particle s.a. extensions of $\hat{Q}^{3}_{phys}(\tau)$. Also, as predicted by Hegerfeldt's paradox, Figure \ref{Fig:Prob2} shows that this violation occurs instantly for any $\tau>0$.

\section{\label{sec:hegr}Time and Position POVM}

In what follows, it will be shown that a single-particle description of the operation $\check{Q}^{0}_{phys}$ is possible only through a POVM approach. Such a result will also allow to explain why the s.a. extensions of $\hat{Q}^{3}_{phys}$ cannot be used directly as localization definitions, the s.a. description needing to be replaced by a description in terms of POVMs.

As stated earlier, the s.a. extensions of $\hat{Q}_{phys}^{0}$ do not define single-particle observables, since they cannot be written as a direct sum of s.a. operators defined to act over the subspaces $\mathcal{H}_{phys}^{\pm}$. Two important consequences of this result are that the perfect temporal localization of a state is only possible outside the single-particle framework and that states with a well defined energy sign will necessarily have a temporal incertitude when described by a classical observer.

The above statements raise the following question: how to describe a measurement associated to the operation $\check{Q}^{0}_{phys}$ in a single-particle framework? To answer it, we observe the fact that the projections of $\hat{Q}^{0}_{phys}(\tau)$ over the subspaces $\mathcal{H}_{phys}^{\pm}$ have one of the deficiency indices equals to zero. This implies that the projected operators $\hat{Q}^{0,\pm}_{phys}(\tau)$ are essentially maximally symmetrical \cite{Gitman2012}, i.e. the closure of these projections leads to maximally symmetrical operators $\underline{\hat{Q}^{0;\pm}_{phys}(\tau)}$  that can be associated with POVMs \cite{Egusquiza2008}. Thus, although a single-particle interpretation cannot be associated to the s.a. extensions. $\tensor[_{\varphi}]{\hat{Q}}{^{0}_{\hspace{-0.1cm}phys}}(\tau)$, the same is not true for the maximally symmetrical operators $\underline{\hat{Q}^{0;\pm}_{phys}(\tau)}$. Therefore, the implementation of the time operator in the proper-time single-particle formalism is given by a POVM and not by an s.a. operator.  In addition, since the boundary conditions associated with the domain of $\underline{\hat{Q}^{0}_{phys}(\tau)}$ do not mix components associated with distinct energy signs, one has that $\underline{\hat{Q}^{0}_{phys}(\tau)} = \underline{\hat{Q}^{0;+}_{phys}(\tau)} \oplus \underline{\hat{Q}^{0;-}_{phys}(\tau)}$ and the domain properties of the operators $\underline{\hat{Q}^{0;\pm}_{phys}(\tau)}$ coincide with those obtained in Section \ref{sec:saq0} for $\underline{\hat{Q}^{0}_{phys}(\tau)}$.

Using Naimark's theorem \cite{Busch1995}, the elements of the POVM associated to $\underline{\hat{Q}^{0;\pm}_{phys}(\tau)}$ can be obtained projecting the eigenfunctions projectors of any s.a. extension $\tensor[_{\varphi}]{\hat{Q}}{^{0}_{\hspace{-0.1cm}phys}}(\tau)$ over $\mathcal{H}^{\pm}_{phys}$. Thus, the set of positive operators $\{\hat{E}_{\tau;\pm}(t)\}$ associated with the POVM defined by $\underline{\hat{Q}^{0;\pm}_{phys}(\tau)}$ is given by
\begin{equation}
\hat{E}_{\tau;\pm}(t) = \sum_{l=0}^{\infty}\sum_{m_{z}=-l}^{l}\ket{\psi^{t,l,m_z}_{\tau;\pm}}\bra{\psi^{t,l,m_z}_{\tau;\pm}}, \nonumber
\end{equation}
with
\begin{equation}
\begin{aligned}
\psi^{t,l,m_z}_{\tau;\pm}(\bs{\pi}) & = \braket{\bs{\pi};\pm|\psi^{t,l,m_z}_{\tau;\pm}} \\ & = \sqrt{\frac{m}{2\uppi}}\frac{Y^{l,m_{z}}(\Omega_{\pi})}{r_{\pi}^{3/2}}\hspace{-0.1cm}\left(\frac{r_{\pi}}{m}\right)^{im\tau}\hspace{-0.1cm}\left(\frac{r_{\pi}}{E_{r_{\pi}} + m}\right)^{\mp imt}.
\end{aligned} \nonumber
\end{equation}
Thus, the probability of finding the system in a time interval $[t_1,t_2]$ for a state described by a density matrix $\rho$ is given by
\begin{equation}
P_{[t_1,t_2]} = \int_{t_{1}}^{t_{2}}dt \mathrm{Tr}(\rho\hat{E}_{\tau;\pm}(t)), \nonumber
\end{equation}
while the completeness relation in $\mathcal{H}_{phys}^{\pm}$ associated with the elements $\hat{E}_{\pm}(t)$ is written as $\int_{-\infty}^{\infty}dt \hat{E}_{\tau;\pm}(t) = \hat{I}^{\pm}_{phys}$, where $\hat{I}^{\pm}_{phys}$ is the identity in $\mathcal{H}_{phys}^{\pm}$. Furthermore, it is important to note that the elements $\hat{E}_{\tau;\pm}(t)$ for different values of $t$ are not orthogonal, since
\begin{widetext}
\begin{equation}
\braket{\psi^{t,l,m_z}_{\tau;\pm}|\psi^{t',l',m'_z}_{\tau;\pm}} = \delta_{ll'}\delta_{m_{z}m_{z}'}\left[\frac{1}{2}\delta(t-t') \pm \frac{1}{2\uppi}\mathrm{P.V.}\left(\frac{i}{t-t'}\right)\right], \nonumber
\end{equation}
\end{widetext}
where $\mathrm{P.V.}$ indicates the principal value. Thus, the strict temporal localization is not possible in the proper-time single-particle formalism and physically acceptable states will necessarily present a temporal uncertainty.

To verify how the imposition of the single-particle character over the $\check{Q}^{0}_{phys}(\tau)$ operation influences the system's spacial localization, the covariance of the quantities $\hat{Q}^{\mu}_{phys}(\tau)$, which follows from the commutation relations \eqref{eq:ASD}, will be taken as a starting point. Denoting by $\hat{U}(\Lambda)$ the unitary representations of the generators $\hat{J}^{\mu\nu}_{phys}$ of the Lorentz group \cite{Tung1985,Taillebois2013}, the relations in \eqref{eq:ASD} imply that, under the action of a arbitrary Lorentz transformation $\Lambda$,
\begin{equation}
\hat{Q}^{\mu'}_{phys}(\tau) \equiv \hat{U}^{\dagger}(\Lambda)\hat{Q}^{\mu}_{phys}(\tau)\hat{U}(\Lambda)= \tensor{\Lambda}{^{\mu}_{\nu}}\hat{Q}^{\nu}_{phys}(\tau). \label{eq:Lor}
\end{equation}
Assuming that $\Lambda$ is a pure boost in the z-axis and $\mu=3$, equation \eqref{eq:Lor} results in
\begin{equation}
\hat{Q}^{3'}_{phys}(\tau) = \tensor{\Lambda}{^{3}_{0}}\hat{Q}^{0}_{phys}(\tau) + \tensor{\Lambda}{^{3}_{3}}\hat{Q}^{3}_{phys}(\tau). \label{eq:Lor1}
\end{equation}
Relationship \eqref{eq:Lor1} implies that, in order to respect the covariance for finite Lorentz transformations, one must have $D_{\hat{Q}^{3}_{phys}(\tau)} = D_{\hat{Q}^{0}_{phys}(\tau)}\cap D_{\hat{Q}^{3}_{phys}(\tau)}$, since the domain of $\hat{Q}^{3}_{phys}(\tau)$ cannot change from one reference frame to another.  However, assuming that in subspace $\mathcal{H}_{phys}^{\xi}$ the time operator is given by $\underline{\hat{Q}^{0,\xi}_{phys}(\tau)}$, there are states in $D_{\tensor[_{\varphi}]{\hat{Q}}{^{3,\xi}_{\hspace{-0.4cm}phys}}(\tau)}$ that will not belong to  $D_{\underline{\hat{Q}^{0,\xi}_{phys}(\tau)}}$, since the boundary condition \eqref{eq:Bound1} imposed by $D_{\tensor[_{\varphi}]{\hat{Q}}{^{3,\xi}_{\hspace{-0.4cm}phys}}(\tau)}$ admits states that don't cancel out faster than $r_{\pi}^{-3/2}$ for $\nu_{\pi}\rightarrow\pm \uppi/2$, as is the case for the eigenstates \eqref{eq:EigenQ3}. Thus, the single-particle character together with finite Lorentz covariance rules out the possibility to use the s.a. extensions of $\hat{Q}_{phys}^{j}(\tau)$ as position operators.

To solve the covariance problem and obtain a spacial localization definition consistent with the finite covariance condition, one can adopt  $\underline{\hat{Q}^{j,\pm}_{phys}(\tau)}$ as definitions of the single-particle position operator in each subspace of well-defined energy sign, since the boundary conditions of these operators are in agreement with those imposed by $\underline{\hat{Q}^{0,\pm}_{phys}(\tau)}$. However, since $\underline{\hat{Q}^{j,\pm}_{phys}(\tau)}$ is symmetrical but not s.a., such a choice will imply in a concept of localization that cannot be associated with projective measures.

In order to define a POVM associated with the operator $\underline{\hat{Q}^{3,\xi}_{phys}(\tau)}$ one observe that $D_{\underline{\hat{Q}^{3,\xi}_{phys}(\tau)}}  \subset D_{\tensor[_{\varphi}]{\hat{Q}}{^{3,\xi}_{\hspace{-0.4cm}phys}}(\tau)}$, i.e. the eigenstates $\mathcal{V}^{z^{n}_{\varphi}}_{\tau;\xi}(\tau)$ of $\tensor[_{\varphi}]{\hat{Q}}{^{3,\xi}_{\hspace{-0.4cm}phys}}(\tau)$ also serves as a basis for $D_{\underline{\hat{Q}^{3,\xi}_{phys}(\tau)}}$. However, unlike for $ D_{\tensor[_{\varphi}]{\hat{Q}}{^{3,\xi}_{\hspace{-0.4cm}phys}}(\tau)}$, that base is improper for $D_{\underline{\hat{Q}^{3,\xi}_{phys}(\tau)}}$, since the states $\mathcal{V}^{z^{n}_{\varphi}}_{\tau;\xi}(\tau)$ do not meet the boundary conditions \eqref{eq:eqh} and, therefore, $\mathcal{V}^{z^{n}_{\varphi}}_{\tau;\xi}(\tau) \not\in D_{\underline{\hat{Q}^{3,\xi}_{phys}(\tau)}}$. Using the improper basis of states $\psi^{z^{n}_{\varphi},\lambda,m_{z}}_{\tau;\xi}(\bs{\pi}) = \mathcal{V}^{z^{n}_{\varphi}}_{\tau;\xi}(\nu_{\pi})\Phi^{m_{z}}(\varphi_{\pi})\mathcal{W}^{\lambda}_{m_z}(\omega_{\pi})$ one has that the operator $\underline{\hat{Q}^{3,\xi}_{phys}(\tau)}$ can be written as
\begin{equation}
\begin{aligned}
&\underline{\hat{Q}^{3,\xi}_{phys}(\tau)} = \sum_{n\in\mathbb{Z}}\sum_{m_{z}\in\mathbb{Z}}\int_{-\infty}^{-1/4}\hspace{-0.6cm}d\lambda z^{n}_{\varphi}\ket{\psi^{z^{n}_{\varphi},\lambda,m_{z}}_{\tau;\xi}}\bra{\psi^{z^{n}_{\varphi},\lambda,m_{z}}_{\tau;\xi}} \\ & = \sum_{n\in\mathbb{Z}}\sum_{m_{z}\in\mathbb{Z}}\int_{-\infty}^{-1/4}\hspace{-0.6cm}d\lambda \left(z^{0}_{\varphi} + \frac{2n}{m}\right)\ket{\psi^{z^{n}_{\varphi},\lambda,m_{z}}_{\tau;\xi}}\bra{\psi^{z^{n}_{\varphi},\lambda,m_{z}}_{\tau;\xi}}, 
\end{aligned} \nonumber
\end{equation}
where \[z^{0}_{\varphi} = \frac{2}{m\uppi}\mathrm{arctan}\left[\frac{\sin\varphi}{1 + \cos\varphi}\tanh\left(\frac{\uppi}{2}\right)\right].\] Since $D_{\underline{\hat{Q}^{3,\xi}_{phys}(\tau)}} \subset D_{\tensor[_{\varphi}]{\hat{Q}}{^{3,\xi}_{\hspace{-0.4cm}phys}}(\tau)}$ for all $\varphi\in(-\uppi,\uppi]$, the above decomposition can be done using the basis of eigenstate of any of the s.a. extensions $\tensor[_{\varphi}]{\hat{Q}}{^{3,\xi}_{\hspace{-0.4cm}phys}}(\tau)$ and, therefore, an integration in $z^{0}_{\varphi}$ allows to rewrite $\underline{\hat{Q}^{3,\xi}_{phys}(\tau)}$ as
\begin{widetext}
\begin{equation}
\begin{aligned}
\underline{\hat{Q}^{3,\xi}_{phys}(\tau)} & = \frac{m}{2}\int_{-1/m}^{1/m}dz^{0}_{\varphi}\sum_{n\in\mathbb{Z}}\sum_{m_{z}\in\mathbb{Z}}\int_{-\infty}^{-1/4}d\lambda \left(z^{0}_{\varphi} + \frac{2n}{m}\right)\ket{\psi^{z^{n}_{\varphi},\lambda,m_{z}}_{\tau;\xi}}\bra{\psi^{z^{n}_{\varphi},\lambda,m_{z}}_{\tau;\xi}} \\
& = \frac{m}{2}\int_{-\infty}^{\infty}dz\sum_{m_{z}\in\mathbb{Z}}\int_{-\infty}^{-1/4}d\lambda z\ket{\psi^{z,\lambda,m_{z}}_{\tau;\xi}}\bra{\psi^{z,\lambda,m_{z}}_{\tau;\xi}}, \label{eq:DecSO}
\end{aligned}
\end{equation}
\end{widetext}
where $\psi^{z,\lambda,m_{z}}_{\tau;\xi}(\bs{\pi}) = \mathcal{V}^{z}_{\tau;\xi}(\nu_{\pi})\Phi^{m_{z}}(\varphi_{\pi})\mathcal{W}^{\lambda}_{m_z}(\omega_{\pi})$ with 
\[\mathcal{V}^{z}_{\tau;\xi}(\nu_{\pi}) = \frac{e^{im\tau\ln(\sec\nu_{\pi})}e^{-im\xi z\nu_{\pi}}}{\uppi^{1/2}(\sec\nu_{\pi})^{3/2}}\]
and $z\in\mathbb{R}$. The decomposition in \eqref{eq:DecSO} allows a continuous description to be retrieved for the system position, although it is non-orthogonal, since
\begin{equation}
\braket{\psi^{z',\lambda',m_{z}'}_{\tau;\xi}|\psi^{z,\lambda,m_{z}}_{\tau;\xi}} = \delta_{m_{z}'m_{z}}\delta(\lambda'-\lambda)\mathrm{sinc}\hspace{-2pt}\left(\frac{m\uppi(z'-z)}{2}\right). \label{eq:NonOrt}
\end{equation}
It is interesting to note that this non-orthogonality decays with $[m(z'-z)]^{-1}$ and, therefore, decreases with the inverse of the number of Compton wavelengths that separate $z'$ from $z$, being relevant only for values of $z'$ very close to $z$.

The above results, along with the fact that the identity $\hat{I}^{\xi}_{phys}$ in $\mathcal{H}^{\xi}_{phys}$ can be written as
\begin{equation}
\hat{I}^{\xi}_{phys} = \frac{m}{2}\int_{-\infty}^{\infty}dz\sum_{m_{z}\in\mathbb{Z}}\int_{-\infty}^{-1/4}d\lambda\ket{\psi^{z,\lambda,m_{z}}_{\tau;\xi}}\bra{\psi^{z,\lambda,m_{z}}_{\tau;\xi}}, \nonumber
\end{equation}
allows to introduce a POVM associated with the operator $\underline{\hat{Q}^{3,\xi}_{phys}(\tau)}$. The positive operators $\{\hat{E}_{\tau;\xi}(z)\}$ associated with that POVM are given by
\begin{equation}
\hat{E}_{\tau;\xi}(z) = \frac{m}{2}\sum_{m_{z}\in\mathbb{Z}}\int_{-\infty}^{-1/4}d\lambda\ket{\psi^{z,\lambda,m_{z}}_{\tau;\xi}}\bra{\psi^{z,\lambda,m_{z}}_{\tau;\xi}}. \label{eq:POVMpos}
\end{equation}
These operators satisfy $\int_{\mathbb{R}}dz\hat{E}_{\tau;\xi}(z) = \hat{I}^{\xi}_{phys}$ and the probability $P_{[z_1,z_2]}(\tau)$ of finding a state of density matrix $\rho$ in a spatial range $[z_1,z_2]$ for a proper-time $\tau$ is given by
\begin{equation}
P_{[z_1,z_2]}(\tau) = \int_{z_{1}}^{z_2}dz \mathrm{Tr}\left(\rho\hat{E}_{\tau;\pm}(z)\right). \nonumber
\end{equation}

To conclude this description of the system's localization in terms of $\underline{\hat{Q}^{3,\xi}_{phys}(\tau)}$ it is necessary to verify that the POVM definition given in \eqref{eq:POVMpos} is not subjected to the causality problems stated by Hegerfeldt in \cite{Hegerfeldt1974,Hegerfeldt1980,Hegerfeldt1985}. Since these results assert the causality violation for strictly localized states as well as exponentially bounded states, it will be shown here that the domain $D_{\underline{\hat{Q}^{3,\pm}_{phys}(\tau)}}$ does not allow such states to be defined.

We will begin with the proof of the nonexistence of strictly localized states. Suppose that the state
\begin{equation}
\ket{\psi} = \int_{-\infty}^{-1/4}d\lambda\alpha(\lambda)\int_{\mathbb{R}}dz \Omega(z)\ket{\psi_{0;+}^{z,\lambda,0}} \nonumber
\end{equation}
is strictly localized at $\tau = 0$. Due to the non-orthogonality given in \eqref{eq:NonOrt}, the strictly localization condition consist in supposing that the density probability
\begin{equation}
\begin{aligned}
\mathcal{P}_{0}(z') & = \mathrm{Tr}(\hat{E}_{0,+}(z')\ket{\psi}\bra{\psi}) \\ & = \frac{m}{2}\Big|\int_{\mathbb{R}}dz \Omega(z)\mathrm{sinc}\left(\frac{m\uppi}{2}(z'-z)\right)\Big|^2,
\end{aligned} \nonumber
\end{equation}
or, in an equivalent way, the probability amplitude
\begin{equation}
p_{0}(z') = \sqrt{\frac{m}{2}}\int_{\mathbb{R}}dz \Omega(z) \mathrm{sinc}\left(\frac{m\uppi}{2}(z'-z)\right), \nonumber
\end{equation}
has a compact support in an interval $\Delta z \subset \mathbb{R}$.

The compact support of $p_{0}(z')$ implies that its Fourier
\begin{equation}
\mathcal{F}_{p_0}(k) = \frac{1}{\sqrt{2\uppi}}\int_{\mathbb{R}}dz' p_0(z') e^{-i\uppi m k z'} \label{eq:transFFF}
\end{equation}
must be analytic in $\mathbb{R}$. Making the change of variables given by $u = m\uppi z$, the Fourier \eqref{eq:transFFF} may be rewritten as
\begin{equation}
\mathcal{F}_{p_0}(k) = \frac{\sqrt{2\uppi}}{(m\uppi)^{3/2}}\mathrm{rect}(k)\mathcal{F}_{\Omega}(k), \label{eq:transFF}
\end{equation}
where
\begin{equation}
\mathrm{rect}(k) = \left\{
\begin{aligned}
0, \quad |k|>\frac{1}{2}  \\
\frac{1}{2}, \quad |k| = \frac{1}{2} \\
1, \quad |k| < \frac{1}{2}
\end{aligned} \right. \nonumber
\end{equation}
is a rectangular function and
\begin{equation}
\mathcal{F}_{\Omega}(k) = \frac{1}{\sqrt{2\uppi}}\int_{\mathbb{R}}du\Omega(u)e^{-iku}. \nonumber
\end{equation}

To verify the conditions imposed by the domain $D_{\underline{Q^{3;+}_{phys}(0)}}$ , the state $\ket{\psi}$ must be written in momentum basis. In this basis
\begin{equation}
\psi_{+}(\bs{\pi}) = \frac{A_{0}}{m\uppi^{3/2}}\frac{\mathcal{F}_{\Omega}(k)}{(\sec\nu_{\pi})^{3/2}}, \nonumber
\end{equation}
where
\begin{equation}
A_{0} = \int_{-\infty}^{-1/4}d\lambda \alpha(\lambda)\mathcal{W}^{\lambda}_{0}(\omega_{\pi}) \nonumber
\end{equation}
and $k = \nu_{\pi}/\uppi$. From the conditions in the domain $D_{\underline{Q^{3;+}_{phys}(0)}}$ it results that $\mathcal{F}_{\Omega}(k)$ must be zero for $k = \pm 1/2$ and must belong to $L^{2}((-1/2,1/2),dk)$, besides being differentiable.

The above properties obtained for $\mathcal{F}_{\Omega}(k)$ imply that the Fourier $\mathcal{F}_{p_0}(k)$ given in \eqref{eq:transFF} must be differentiable and have compact support in $[-1/2,1/2]\subset\mathbb{R}$, since $\mathcal{F}_{p_0}(k) \propto \mathrm{rect}(k)\mathcal{F}_{\Omega}(k)$. Therefore, the function $\mathcal{F}_{p_0}(k)$ cannot be analytic in $\mathbb{R}$ and $p_{0}(z')$ cannot have compact support, which demonstrates the nonexistence of strictly localized states with respect to the localization definition associated to $\underline{\hat{Q}^{3,\xi}_{phys}(\tau)}$.

The demonstration of the nonexistence of states that are compatible with $\underline{\hat{Q}^{3,\xi}_{phys}(\tau)}$ and have a probability amplitude $p_0(z')$ with tails bounded by an exponential decay $e^{-A|z|}$ follows the same reasoning presented above. In this case, the exponential behavior of the tails of $p_0(z')$ would imply the analyticity of the Fourier $\mathcal{F}_{p_0}(k)$ for $|\mathrm{Im}(k)| < A$. However, the domain $D_{\underline{Q^{3; +}_{phys}(0)}}$ implies that over $\mathbb{R}$ the function $\mathcal{F}_{p_0}(k)$ must have compact support in $[-1/2,1/2]$ and therefore the condition of analyticity in $|\mathrm{Im}(k)| < A$ cannot be satisfied, which concludes the demonstration of the nonexistence of states with bounded exponential decay.

\section{\label{sec:disc}Discussion}

The problem of localization in RQM was approached through a single-particle formalism parameterized by the system's proper-time. This parameterization has a fundamental character since it does not depend on the properties of an external observer and, therefore, corresponds to an intrinsic approach to the problem of localization in RQM. Physically, adopting the proper-time gauge as fundamental amounts to state that the system's time would be observed as classical only if it were possible to define an observer from a comoving quantum frame as those proposed in \cite{Giacomini2019n}.

Although the quantum operations $\check{Q}^{\mu}_{phys}(\tau)$ may be associated with s.a. operators $\tensor[_{\varphi}]{\hat{Q}}{^{\mu}_{\hspace{-0.1cm}phys}}(\tau)$, it has been shown that these s.a. extensions cannot be associated with a single-particle description of the system, a result that follows from the nonexistence of single-particle s.a. extension of $\hat{Q}^{0}_{phys}(\tau)$ and from the four-vector character of the operations $\check{Q}^{\mu}_{phys}(\tau)$, that ensures the covariance of the definition. The immediate consequence of this result for systems with a well-defined energy sign is the necessity of introducing a quantum description for the variables $Q^{0}_{phys}(\tau)$ based on POVMs connected to the maximally symmetrical operators $\underline{Q^{0;\pm}_{phys}(\tau)}$, which implies the impossibility of constructing temporally localized states in $\mathcal{H}^{\pm}_{phys}$, i.e. physically acceptable single-particle states have an intrinsic temporal uncertainty.

For the spatial variables $Q^{j}(\tau)$, it was shown that certain parameterizations of the s.a. extensions $\tensor[_{\varphi}]{\hat{Q}}{^{j}_{\hspace{-0.1cm}phys}}(\tau)$ may correspond to a single-particle description of the system's position. However, the four-vector character of the variables $Q^{\mu}(\tau)$ imposes restrictions on the quantum description that cannot be satisfied by the operators  $\tensor[_{\varphi}]{\hat{Q}}{^{j}_{\hspace{-0.1cm}phys}}(\tau)$ if $\underline{Q^{0;\xi}_{phys}(\tau)}$ is adopted as the temporal component of the operator associated with the system's four-position. Consequently, as happened with $Q^{0}_{phys}(\tau)$, the variables $Q^{j}_{phys}(\tau)$ also need to be described by POVMs associated to the closed operator $\underline{Q^{j;\xi}_{phys}(\tau)}$. This description is not subject to the predictions concerning Hegerfeldt's paradox, since the domain of $\underline{Q^{j;\xi}_{phys}(\tau)}$ implies the impossibility of strict localization as well as states with bounded exponential tails. It is important to emphasize the crucial role played by the domain analysis in the presented results. This demonstrates the relevance of going beyond the algebraic properties of operators in RQM, since the particular characteristics of unbounded operators may become of paramount importance.

It is worth noting that the impossibility of defining strictly localized states in a single-particle approach is in agreement with the idea that such localization would involve energies that would lead to a regime in which the phenomena of creation and annihilation of particles could no longer be disregarded. As obtained in \eqref{eq:NonOrt}, the distance between two orthogonal position in the z-axis is at least of two Compton wavelength, a range that is in agreement with what is expected from a regime with fixed number of particles as RQM.

Finally, it is necessary to emphasize the need to understand how the present proposal connects itself to measurements parameterized by classical observer parameters, as well as the consequences of temporal uncertainty for these measurements. This connection is important to discuss causality in contexts where measurements are parameterized by quantities as the observer time and will be the subject of a future work. However, it should be noted that, due to the intrinsic temporal uncertainty of quantum states, any measurement parameterized by a classical observer will correspond to a coarse-grained description of the system.

For the sake of completeness, it should be stressed that in \cite{VonZuben2000} the Dirac Hamiltonian formalism was applied in a different way to a free spinless massive particle to introduce a new perspective concerning Hegerfeldt's paradox. However, a mathematical inaccuracy pointed by the present authors in \cite{Taillebois2020c} is responsible for invalidating those results.

\vspace{\baselineskip}
\begin{acknowledgments}
We are thankful for the support provided by Brazilian agencies CAPES (PROCAD2013), CNPq (\#459339/2014-1, \#312723/2018-0), FAPEG (PRONEX \#201710267000503, PRONEM \#201710267000540) and the Instituto Nacional de Ciência e Tecnologia - Informação Quântica (INCT-IQ).
\end{acknowledgments}

\appendix

\section{\label{app:CSCO} Properties of the  \texorpdfstring{$\hat{Q}^{3}_{phys}(\tau)$}{Q3phys} CSCO}

The full CSCO related to $\hat{Q}^{3}_{phys}(\tau)$ is formed by the operators $\hat{J}^{12}_{phys}$ and $\hat{O}_{phys} \equiv (\hat{J}^{12}_{phys})^2 - (\hat{J}^{01}_{phys})^2 - (\hat{J}^{02}_{phys})^2$. While the s.a. character of $\hat{J}^{12}_{phys}$ defined over its natural domain with a periodicity restriction in $\varphi_{\pi} \in [0,2\uppi)$ is easily verifiable and follows the usual non-relativistic approach, operator $\hat{O}_{phys}$ needs a more detailed treatment. 

Adopting hyperbolic coordinates and the separation of variables $\psi(\omega_{\pi},\varphi_{\pi}) = \mathcal{W}^{\lambda}_{m_{z}}(\omega_{\pi})\Phi^{m_{z}}(\varphi_{\pi})$ for the differential equations $\check{O}_{phys}\psi(\omega_{\pi},\varphi_{\pi}) = \lambda\psi(\omega_{\pi},\varphi_{\pi})$ and $\check{J}^{12}\psi(\omega_{\pi},\varphi_{\pi}) = m_{z}\psi(\omega_{\pi},\varphi_{\pi})$, one has that
\begin{eqnarray*}
-i\frac{d}{d\varphi_{\pi}}\Phi^{m_{z}}(\varphi_{\pi}) & = & m_{z}\Phi^{m_{z}}(\varphi_{\pi}), \\
\left(\frac{d^2}{d\omega_{\pi}^2} + \coth(\omega_{\pi})\frac{d}{d\omega_{\pi}} - \lambda\right)\mathcal{W}^{\lambda}_{m_{z}}(\omega_{\pi}) & = &  \frac{m_{z}^2\;\mathcal{W}^{\lambda}_{m_{z}}(\omega_{\pi})}{\sinh^{2}(\omega_{\pi})},
\end{eqnarray*}
which leads to $\Phi^{m_{z}}(\varphi_{\pi}) = e^{im_{z}\varphi_{\pi}}/\sqrt{2\uppi}$, with $m_{z} \in \mathbb{Z}$, and to the differential equation $\check{O}^{m_{z}}_{phys}\mathcal{W}^{\lambda}_{m_{z}}(\omega_{\pi}) = \lambda\mathcal{W}^{\lambda}_{m_{z}}(\omega_{\pi})$, where
\begin{equation}
\check{O}_{phys}^{m_{z}} \equiv \frac{d^{2}}{d \omega_{\pi}^{2}} + \coth(\omega_{\pi})\frac{d}{d\omega_{\pi}} - \frac{m_{z}^{2}}{\sinh^{2}(\omega_{\pi})}. \label{eq:opO}
\end{equation}

Starting with $C_{0}^{\infty}(0,\infty)$ as test function space and taking into account the fact that $\sinh(\omega_{\pi})$ is locally a.c. in $[0,\infty)$, it results that operation \eqref{eq:opO} is Lagrange s.a. with regard to the measure $d\mu(\omega_{\pi}) = m^{3}\sinh(\omega_{\pi})d\omega_{\pi}$. Thus, choosing $C_{0}^{\infty}(0,\infty)$ as initial domain for the operator $\hat{O}_{phys}^{m_{z}}$, its adjoint is defined by the acting rule \eqref{eq:opO} over the domain
\begin{widetext}
\begin{equation}
D_{\hat{O}_{phys}^{m_{z}}}^{*} = \left\{ \phi_{*}(\omega_{\pi}) \Big| \phi_{*}(\omega_{\pi}), \frac{d \phi_{*}(\omega_{\pi})}{d\omega_{\pi}} \text{ are a.c. in } [0,\infty)
\text{ and } \phi_{*}(\omega_{\pi}), \check{O}_{phys}^{m_{z}}\phi_{*}(\omega_{\pi}) \in L^{2}([0,\infty), d\mu(\omega_{\pi}))\right\}. \label{eq:DOMO}
\end{equation} 
\end{widetext}

To verify the s.a. character of the adjoint operator $\hat{O}_{phys}^{m_{z}*}$ it is helpful to start with the change of variables $u_{\pi} = \cosh(\omega_{\pi})$ ($u_{\pi} \in [1,\infty)$), which allows to rewrite $\check{O}^{m_{z}}_{phys}\mathcal{W}^{\lambda}_{m_{z}}(\omega_{\pi}) = \lambda\mathcal{W}^{\lambda}_{m_{z}}(\omega_{\pi})$ as the associated Legendre differential equation
\begin{widetext}
\begin{equation}
\left((u^2_{\pi} -1)\frac{d^2}{du^2_{\pi}} + 2u_{\pi}\frac{d}{du_{\pi}} - \frac{m_{z}^2}{u^2_{\pi} -1}\right)\mathcal{W}^{\lambda}_{m_{z}}(u_{\pi}) = \upnu(\upnu+1)\mathcal{W}^{\lambda}_{m_{z}}(u_{\pi}), \label{eq:DifLag}
\end{equation}
\end{widetext}
where $\upnu = -\frac{1}{2}+i\Lambda(\lambda)$, with$\Lambda(\lambda) = \sqrt{-\frac{1}{4}-\lambda}$.

Using the variable $u_{\pi}$, the absolute continuity condition from domain \eqref{eq:DOMO} implies that  $\sqrt{u_{\pi}^2 - 1}\frac{d\phi_{*}(u_{\pi})}{du_{\pi}}$ must be a.c. for $u_{\pi}\in [1,\infty)$. Then, this result implies that $\sqrt{u_{\pi}^2 - 1}\frac{d\phi_{*}(u_{\pi})}{du_{\pi}}$ must be limited at the closed end $u_{\pi} = 1$ and, therefore,
\begin{equation}
\lim_{u_{\pi}\rightarrow 1}(u_{\pi}^{2} - 1)\frac{d\phi_{*}(u_{\pi})}{du_{\pi}} = 0, \quad \forall \; \phi_{*} \in D^{*}_{\hat{Q}^{m_{z}}_{phys}}. \label{eq:COnasd}
\end{equation}
However, it is known from literature \cite{Littlejohn2011,Sousa2018} that, provided with the boundary condition \eqref{eq:COnasd}, the associated Legendre differential equation \eqref{eq:DifLag} defines a s.a. operator. Thus, it follows that the operator $\hat{O}^{m_{z}*}_{phys}$ is in fact s.a. and $\hat{O}^{m_{z}}_{phys}$ is essentially s.a. (deficiency indices $\eta=(0.0)$), its only s.a. extension being the adjoint operator $\hat{O}^{m_{z}*}_{phys} = \underline{\hat{O}^{m_{z}}_{phys}}$.

The spectrum and eigenfunctions of the s.a. operator given by the associated Legendre differential equation with boundary condition \eqref{eq:COnasd} are known from the literature \cite{Littlejohn2011,Sousa2018} and, in the present case, are given by $\lambda \in (-\infty,-\frac{1}{4}]$ and
\begin{equation}
\mathcal{W}^{\lambda}_{m_{z}}(\omega_{\pi}) = \mathcal{N}^{|m_{z}|}_{\lambda}P^{-|m_{z}|}_{-\frac{1}{2} + i\Lambda(\lambda)}(\cosh\omega_{\pi}), \nonumber
\end{equation}
where $P^{-|m_{z}|}_{-\frac{1}{2} + i\Lambda(\lambda)}(\cosh\omega_{\pi})$ denotes the associated conical function (associated Legendre function of the first kind) and
\begin{equation}
\mathcal{N}^{|m_{z}|}_{\lambda} = \frac{|\Gamma(\frac{1}{2} + |m_{z}| + i\Lambda(\lambda))|}{m^{3/2}}\sqrt{\frac{\sinh(\uppi\Lambda(\lambda))}{2\uppi}}. \nonumber
\end{equation}

The results presented above, together with those obtained for the single-particle s.a. extensions $\tensor[_{\varphi}]{\hat{Q}}{^{3,\xi}_{\hspace{-0.4cm}phys}}(\tau)$, allow to write the eigenfunctions of the CSCO of $\tensor[_{\varphi}]{\hat{Q}}{^{3,\xi}_{\hspace{-0.4cm}phys}}(\tau)$ as in \eqref{eq:EigenQ3}, the orthogonality and completeness relations presented in \eqref{eq:Iju} being a consequence of the relationships
\begin{widetext}
\begin{eqnarray*}
\int_{0}^{\infty}\sinh(\omega_{\pi})d\omega_{\pi}P^{-|m_{z}|}_{-\frac{1}{2} + i\Lambda'}(\cosh\omega_{\pi})P^{-|m_{z}|}_{-\frac{1}{2} + i\Lambda}(\cosh\omega_{\pi}) & = & \frac{\uppi \delta(\Lambda'-\Lambda)}{\Lambda \sinh(\uppi\Lambda)|\Gamma(\frac{1}{2} + |m_{z}| + i\Lambda)|^2}, \\
\int_{0}^{\infty}d\Lambda \frac{\Lambda\sinh(\uppi\Lambda)}{\uppi}\big|\Gamma\left(\frac{1}{2} + |m_{z}| + i\Lambda\right)\big|^2 P^{-|m_{z}|}_{-\frac{1}{2} + i\Lambda}(\cosh\omega_{\pi}')P^{-|m_{z}|}_{-\frac{1}{2} + i\Lambda}(\cosh\omega_{\pi}) & = & \delta(\cosh\omega_{\pi}' - \cosh\omega_{\pi}).
\end{eqnarray*}
\end{widetext}

\bibliography{bib/Bibliografia,bib/bibtiqr,bib/Livros}

\end{document}